\documentclass{ws-ijmpa}
\usepackage{amssymb,rotating}
\newcommand{\zz}{{\mathbb Z}}
\newcommand{\re}{{\mathbb R}}
\newcommand{\cc}{{\mathbb C}}
\newcommand{\moduli}{\mathcal{M}}
\newcommand{\lagr}{\mathcal{L}}
\newcommand{\bsp}{\mathcal{B}^*}
\newcommand{\obsv}{\mathcal{O}}
\newcommand{\DX}{\mathcal{D}X}
\newcommand{\Tr}{\mathop{\rm Tr}\nolimits}
\newcommand{\im}{\mathop{\rm Im}\nolimits}
\newcommand{\spin}{\mathop{\rm Spin}\nolimits}
\begin{document}

\markboth{Kevin Iga}
{What do Topologists want from Seiberg--Witten theory?}

\catchline{}{}{}

\title{WHAT DO TOPOLOGISTS WANT\\
FROM SEIBERG--WITTEN THEORY?}

\author{\footnotesize KEVIN IGA}

\address{Natural Science Division\\
Pepperdine University\\
24255 Pacific Coast Hwy.\\
Malibu, CA 90263, USA}

\maketitle

\pub{Received (Day Month Year)}{Revised (Day Month Year)}

\begin{abstract}
In 1983, Donaldson shocked the topology world by using instantons from
physics to prove new theorems about four-dimensional manifolds, and he
developed new topological invariants.  In 1988, Witten showed how these
invariants could be obtained by correlation functions for a twisted
$N=2$ SUSY gauge theory.  In 1994, Seiberg and Witten discovered dualities
for such theories, and in particular, developed a new way of looking at
four-dimensional manifolds that turns out to be easier, and is conjectured
to be equivalent to, Donaldson theory.

This review describes the development of this mathematical subject, and
shows how the physics played a pivotal role in the current understanding
of this area of topology.
\end{abstract}

When, in 1994, Nathan Seiberg and Edward Witten introduced
Seiberg--Witten theory to the physics world, the study of
supersymmetry was revolutionized.  But surprisingly, mathematicians
were also getting excited about Seiberg--Witten theory.  They started
asking physicists questions and seeking answers.  They started making new
discoveries in their mathematical fields, supposedly because of the
physics.  And many of these mathematicians were not mathematical
physicists, but topologists.  What did topologists want from
Seiberg--Witten theory?

Stereotypically, a physicist supposedly comes up with a
mathematical way of phrasing a physics problem, and a mathematician
helps solve the problem.  The result is answers for the physicist, and
interesting areas of research for the mathematician.  For Seiberg--Witten
theory, however, the situation was reversed.  Mathematicians were looking
to the physicists for answers to mathematical questions.

There were actually many cases of this sort of interaction, especially
in the last twenty years.  Developments in high-energy physics have
led to completely new techniques in topology, resulting in an
impressive growth in our knowledge of four-dimensional manifolds,
complex manifolds, knot theory, differential geometry, and symplectic
geometry.

To understand what Seiberg--Witten theory has to do with topology,
though, we need to delve into one of these earlier cases: in the early
1980s, when physics affected four-dimensional topology through the
study of instantons.  In 1983, Simon Donaldson showed how studying
instantons led to new theorems and powerful new techniques in
understanding the topology of four-dimensional manifolds.  Both the
instanton revolution of 1983 and the Seiberg--Witten revolution of
1994 are interesting cases of physics leading to new breakthroughs in
mathematics.

The purpose of this review is to study these two breakthroughs and
examine how questions from physics helped address fundamental
questions in four-dimensional topology.  We will also see the impact
of physics on what was known in the subject, and how the physics led
to the discovery of intriguing relationships to other areas of mathematics.
As a mathematician, I hope to convey to a physics audience an
appreciation what has happened, in language that I hope is as
physics-friendly as possible.\footnote{This review is based primarily
on a lecture I gave to the Stanford physics department in 1998, but
more details taken from a talk I gave to the UCLA mathematics
department in 2001, though the mathematics talk is translated into
physics language for the purposes of this review, as far as I was
able.}

This article will outline the development of the subject in roughly
chronological order.  In sections 1 and 2, I will describe the problem
of classifying four-dimensional manifolds, and indicate what was known
before physics got involved.  In sections 3 through 9, I will describe
S. Donaldson's work in 1983 showing how finding $SU(2)$ instantons on
the manifold can help solve some of these questions, while introducing
new topological invariants.  Section 10 describes Witten's 1988
derivation of these topological invariants using a supersymmetric
topological quantum field theory.  Sections 11 through 13 describe how
notions of duality in supersymmetry discovered by Seiberg and Witten
in 1994 give rise to a new topological quantum field theory that is
easier to handle mathematically.  Sections 15 through 17 describe some
advances in four-dimensional topology using this dual theory.  Section 18
has some philosophical conclusions.

\section{Topology}
\begin{figure}
\begin{center}
\begin{tabular}{ccccc}
\rule{0in}{1in}
\epsfig{file=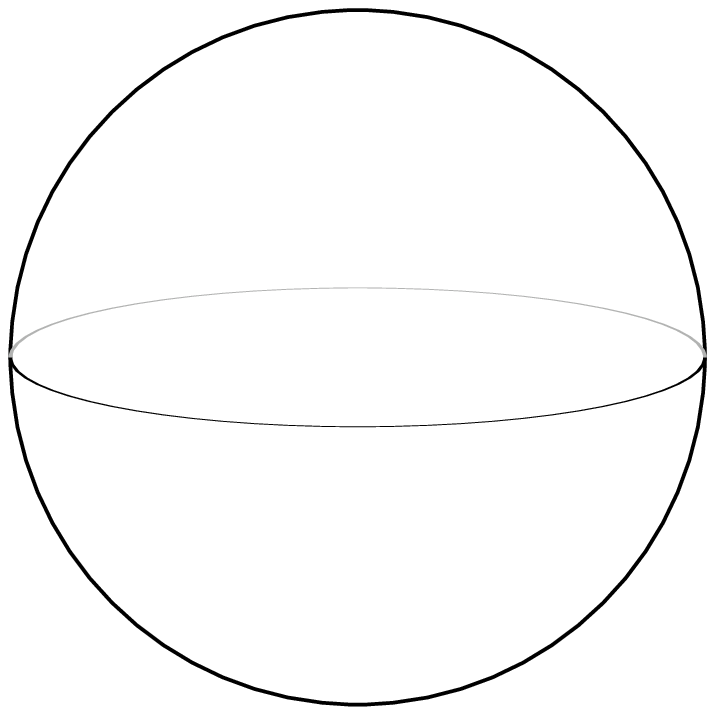,width=.8in,bbllx=200,bblly=200,bburx=400,bbury=400}
&
\epsfig{file=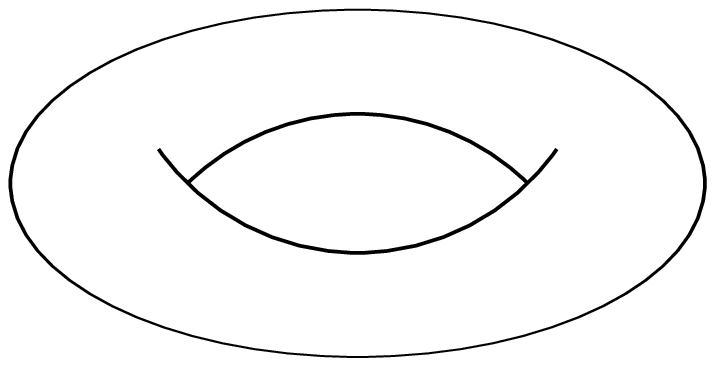,width=.8in,bbllx=200,bblly=200,bburx=400,bbury=350}
&
\epsfig{file=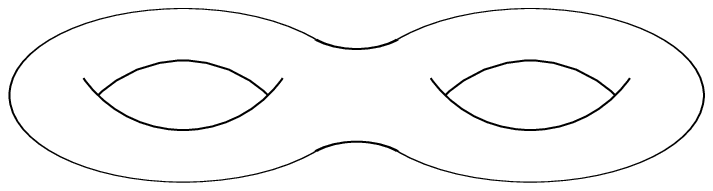,width=.8in,bbllx=200,bblly=200,bburx=400,bbury=350}
&
\epsfig{file=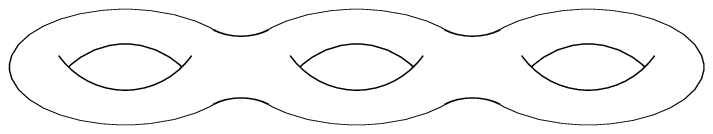,width=.8in,bbllx=200,bblly=200,bburx=400,bbury=350}
&\raisebox{.5in}{$\cdots$}\\  
$\chi(S^2)=2$
&$\chi(T^2)=0$
&$\chi(T^2\#T^2)=-2$
&$\chi(T^2\#T^2\#T^2)=-4$
&$\cdots$\\
\rule{0in}{1in}
\epsfig{file=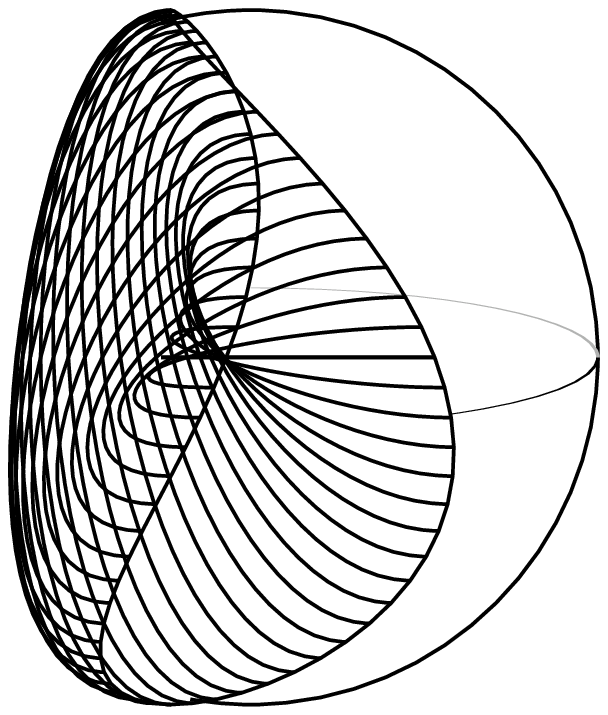,width=.8in,bbllx=200,bblly=200,bburx=400,bbury=400}
&
\epsfig{file=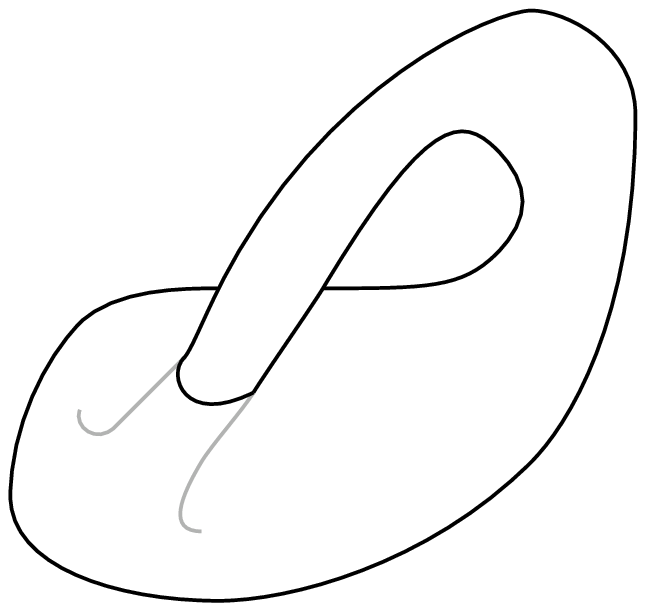,width=.8in,bbllx=200,bblly=200,bburx=400,bbury=400}
&
\epsfig{file=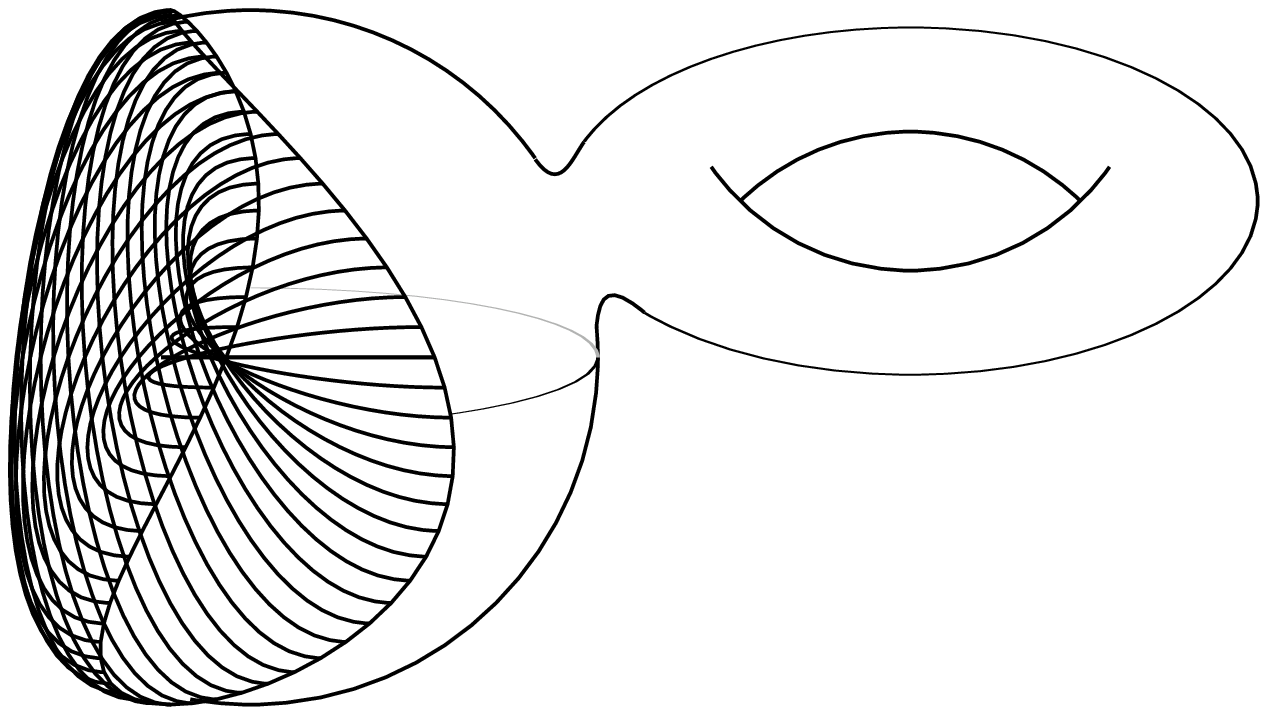,width=.8in,bbllx=200,bblly=200,bburx=590,bbury=400}
&
\epsfig{file=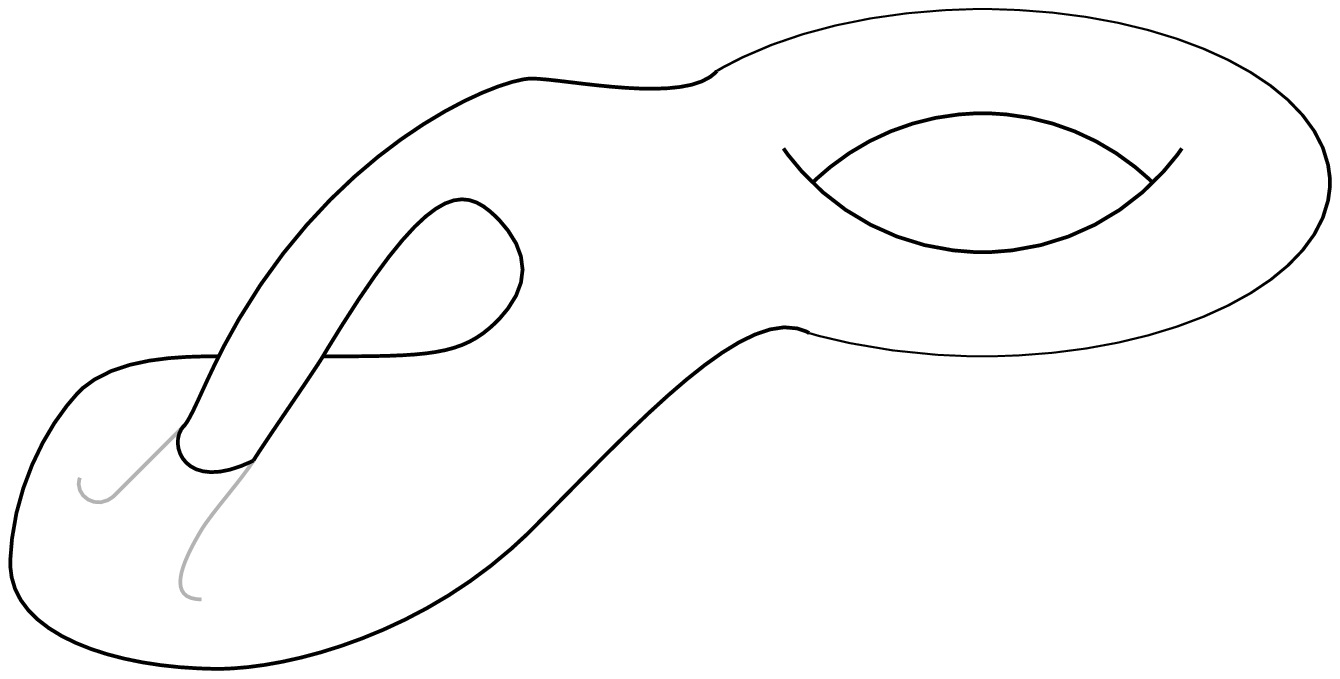,width=.8in,bbllx=200,bblly=200,bburx=580,bbury=400}
&\raisebox{.5in}{$\cdots$}\\  
$\chi(\re P^2)=1$
&$\chi(K)=0$
&$\chi(\re P^2\# T^2)=-1$
&$\chi(K\# T^2)=-2$
&$\cdots$\\
\end{tabular}
\end{center}
\caption{Classification of compact connected surfaces
without boundary: orientable surfaces are on top and non-orientable surfaces
are on the bottom.  The non-orientable ones cannot be
embedded in $\re^3$ so those drawings are to be suggestive at best.
The Euler characteristic $\chi$ of each is shown beneath each picture.}
\label{fig:surfaces}
\end{figure}

One of the central problems of topology is to classify manifolds.  Two
manifolds are said to be the same if there is a diffeomorphism between
them.  To illustrate this problem, consider the classification of
compact connected surfaces without boundary.  This problem was solved
by Poincar\'e in the early twentieth century, and it goes like this:
some surfaces are orientable, and some are not.  Here is a list of the
compact connected surfaces without boundary that are orientable: the
two-sphere $S^2$, the two-torus $T^2$, the double-torus (like a torus
but with two handles), the triple-torus, and so on (see
Figure~\ref{fig:surfaces}).  A good way to think of these is as a
connected sum of tori.  The {\em connected sum} of two connected
surfaces $X$ and $Y$ is what you get when you remove a disk from $X$
and a disk from $Y$, then glue the result along the boundary.  This
connected sum is written $X\# Y$.  Then the double torus is
$T^2\#T^2$, the triple-torus is $T^2\#T^2\#T^2$, and so on.
The non-orientable compact connected surfaces without boundary are:
the real projective plane $\re P^2$, the Klein bottle $K$, $\re
P^2\#T^2$ (which is the same thing as $\re P^2\#K$), and so on.

The Euler characteristic $\chi$, defined in a standard course in
algebraic topology, is a number that is easy to assign to each
surface.  Once you know whether or not a surface is orientable, the
Euler characteristic uniquely determines the surface.

This is what is meant by a classification of compact connected
surfaces without boundary.  More generally, we would like to classify
manifolds, the $n$-dimensional version of surfaces.  Whether or not we
insist on connectedness is not very important, since any disconnected
manifold is just a union of connected manifolds.  The criterion of
compactness is more worthwhile, since any open subset of a manifold is also
a manifold, and we don't want to get bogged down in the classification
of open subsets.  There is more here than that, and much of it is very
interesting, but as we will see, it is a hard enough question to
classify compact manifolds, that it makes sense not to be too ambitious
too quickly.  For similar reasons we will focus on manifolds without boundary.
From now on, when I mention classification of manifolds,
I will mean the classification of compact manifolds without boundary.

If we were to pattern the project of classification of manifolds after
the classification above for surfaces, then one way to describe the
problem would be to say that we wish to assign some mathematical
object (such as a number, a group, or anything just as easy to
understand) to each manifold (hopefully in a way that is easy to
compute) so that if two manifolds are diffeomorphic, they have the
same mathematical object (in which case the object is called
a topological invariant), and so that if two manifolds are not diffeomorphic,
then they are not assigned the same mathematical object (in which case
the topological invariant is called a complete topological invariant).

In the case of surfaces, we had two important topological invariants:
the Euler characteristic (a number assigned to each surface), and the
orientability (a ``yes'' or a ``no'' assigned to each surface,
answering the question of whether or not it was orientable).  Neither
alone is a complete topological invariant of compact connected
surfaces without boundary; but the ordered pair is.

In general, we don't hope to come up with a single object that is our
complete topological invariant right away; we expect to come up with
many topological invariants which together (we hope) classify manifolds
completely.

The classification of two-dimensional manifolds is mentioned above
(the classification of surfaces).  The classification of
one-dimensional (compact, connected, no boundary) manifolds is also
easy.  There is only one: the circle $S^1$.  If we allowed for
non-compact manifolds we could have the real line $\re$, and if we
allowed for boundary we could have intervals like $[0, 1]$.  Similarly
for zero-dimensional compact connected manifolds: the only such
manifold is a single point.

Now that we have the easy examples out of the way, we might ask about
$n$-dimensional manifolds where $n\ge 3$.  There is much that is known
and much that is not known for such dimensions.  Throughout the 1930s
through the 1950s, the subject of algebraic topology developed.
Algebraic topology defined many kinds of topological invariants that
were defined for $n$-dimensional manifolds (in fact they were usually
defined for arbitrary topological spaces).  For instance if $X$ is a
connected space, its fundamental group $\pi_1(X)$ is a group, and if
two manifolds are diffeomorphic, then they have the same fundamental
group.  Therefore, $\pi_1$ is a topological invariant.

There are generalizations $\pi_2(X), \pi_3(X), \dots$, that are also
topological invariants, which are actually abelian groups.  There are
other sequences of topological invariants that are groups: the
homology of a manifold $X$ is a sequence of abelian groups $H_0(X),
H_1(X), \dots$, and the cohomology $H^0(X), H^1(X), \dots$, and there
are others.  A brief account for physicists is found in Nash and Sen's
book {\em Topology and Geometry for Physicists}\cite{NS}, and a more
complete text on the subject is {\em Elements of Algebraic Topology}
by J. Munkres\cite{Mu}.

For compact manifolds, these groups are all finitely generated, and the
point is that inasmuch as finitely generated groups are understood
(they are not) and inasmuch as finitely generated abelian groups are
understood (they are), these invariants should make it easier to
understand the problem of classification of manifolds.  The problem is
that it is not clear whether or not these form a set of complete
invariants, and furthermore, which values of the invariants are
possible.

Actually, it is possible to prove that in dimension 4 and higher, any
group with finitely many generators and relations can be $\pi_1(X)$
for some manifold $X$.  This can be done explicitly enough that the
classification of manifolds would also produce a classification of
groups with finitely many generators and relations.  The bad news is
that the classification of groups with finitely many generators and
relations has been proven to be impossible,\cite{B,BHP} and
therefore, the classification of manifolds must be impossible, too.

This would seem to answer the main problem in a spectacularly negative
fashion: if $n\ge 4$, then the classification of compact $n$-manifolds
without boundary is algorithmically impossible.

But this is not the end of the story.  We could restrict our attention
to simply connected manifolds (those for which $\pi_1(X)$ is the
trivial group), or manifolds with $\pi_1(X)$ some group that is easy
to understand (finite groups, cyclic groups, etc.).  And it is
precisely for dimensions 4 and higher that we know of many, many
manifolds that are simply connected, so classifying simply connected
manifolds (as before, compact, connected, no boundary) is a very
interesting question, and perhaps one we can hope to answer.

For example, in dimension 4, we have the sphere $S^4$, we have
$S^2\times S^2$, we have the complex projective plane $\cc P^2$, there
are connected sums of these\footnote{The connected sum $X\#Y$ of
$n$-dimensional manifolds $X$ and $Y$ is defined analogously to
surfaces: remove a small ball (a copy of $B^n$) from $X$ and from $Y$,
and glue along their boundary (a copy of $S^{n-1}$).  For dimension
three and higher, $X\#Y$ is simply connected if and only if $X$ and
$Y$ are.}, and there are many more that arise naturally in algebraic
geometry.

In dimensions 5 and higher, remarkably, the problem of classifying
simply connected compact manifolds without boundary is solved, whereas
the analogous classification in dimensions 3 and 4 is still unsolved
today.  This strange circumstance, suggesting that dimensions 5 and
higher are easier than dimensions 3 and 4, comes about because there
are certain techniques that are very powerful, but require a certain
amount of room before you can use them.  A readable account can be
found in Kosinski's {\em Differential Manifolds}\cite{Ko} and
Ranicki's book on surgery theory\cite{Ra} (which should be read in
that order).  This classification also extends to the
classification of manifolds whose $\pi_1(X)$ is understood
sufficiently well.

This leaves the problem of classifying manifolds of dimensions 3 and
4.  Again, for dimension four, we would like to insist that the
manifolds are simply connected, or at least that $\pi_1(X)$ be
sufficiently well understood.  In dimension 3, it is not clear whether
or not we need to be concerned with $\pi_1$, and it is not known if
there are other simply connected compact three-dimensional manifolds
without boundary other than the three-sphere $S^3$.  It is interesting
that the classification problem is not solved in dimensions 3 and 4,
the two dimensions that have long been of interest to physics (space
and space-time) until the advent of string theory.  I will focus on
dimension four, since that is where Seiberg--Witten theory has had its
impact.

\section{What was classically known in dimension four}
Before the 1980s, there was not much known about simply connected
four-dimensional manifolds.  It was possible to compute homology and
cohomology groups, and so on, but invariants like these from algebraic
topology gave limited information, and it was not clear whether or not
there was more to the classification story.

The homology groups look like this:
\begin{eqnarray*}
H_0(X^4)&\cong& \zz\\
H_1(X^4)&\cong& 0\\
H_2(X^4)&\cong& \zz^{b_2}\\
H_3(X^4)&\cong& 0\\
H_4(X^4)&\cong& \zz
\end{eqnarray*}
and the higher homology groups are all trivial.  The vanishing of
$H_1$ occurs because $X^4$ is simply connected (using the Hurewicz
theorem); the vanishing of $H_3$ occurs because of Poincar\'e duality.
So if you were to use only homology, the only topological invariant we
could get was one number: the second Betti number $b_2$.

The cohomology groups can be calculated using the universal coefficient
theorem, and in this case, the table for cohomology groups is identical to
the one for homology groups above.  But the cohomology
groups have some extra information, because cohomology classes can be
multiplied via the wedge product.  In our case, the only case to
consider is multiplying two elements of $H^2(X^4)$, which gives
rise to an element of $H^4(X^4)$.  We can view this as number by
integrating over $X^4$:
\begin{equation}
I(\omega_1,\omega_2)=\int_{X^4} \omega_1 \wedge \omega_2
\label{eqn:wedge}
\end{equation}
where $\omega_1$ and $\omega_2$ are elements of $H^2(X^4)$.  This can
be viewed as a bilinear form on $H^2(X^4)$, taking two cohomology classes
and returning a number.

If we use singular cohomology with integer coefficients instead of
using differential forms, it would be more apparent that
(\ref{eqn:wedge}) is an integer, and in that language, the wedge
product is called the cup product.\cite{Mu}

By Poincar\'e duality, we can interpret (\ref{eqn:wedge}) in terms of
homology instead of cohomology, and this is what happens: an element
of $H_2(X^4)$ can be viewed as a surface embedded in $X^4$, and if
$\Sigma_1$ and $\Sigma_2$ are two such, they will generically
intersect in a finite set of points.  If these are counted with
appropriate signs, the number of points in the intersection will be
the integer that corresponds to (\ref{eqn:wedge}).  In this way, we
can define $I(\Sigma_1,\Sigma_2)$ as a bilinear form on
$H_2(X^4)$.\cite{Ko}

Whichever way you wish to think of it, there is a bilinear form on
$H_2(X^4)$ or equivalently on $H^2(X^4)$ called the intersection form,
and it is symmetric, integer-valued, and non-degenerate.  If we choose
a basis for $H_2(X^4)$, this intersection form can be viewed as a
square $b_2\times b_2$ matrix of integers.  This matrix is symmetric
and its determinant is $\pm 1$.

This intersection form is a topological invariant: every simply
connected compact four-dimensional manifold without boundary gives
rise to an integer-valued $b_2\times b_2$ symmetric matrix with
determinant $\pm 1$.  This would be convenient (matrices are very
convenient and easy to understand), except for one problem: the
intersection form may be a bilinear form on $H_2(X^4)$; but
identifying it as a matrix requires choosing a basis.  Changing this
basis with an invertible integer-valued matrix $S$ will change the
matrix $A$ to $S^TAS$, and both $A$ and $S^TAS$ are descriptions of
the same bilinear form but in different bases.

The classification of symmetric integer-valued bilinear forms with determinant
$\pm 1$, up to integer change of basis, is a difficult subject in general.
If we were to allow any real change of basis, the classification of these
bilinear forms is just a matter of counting the number of positive eigenvalues
and the number of negative eigenvalues (since the determinant is $\pm 1$,
there are no zero eigenvalues).  Let $b_2^+$ be the number of positive
eigenvalues and $b_2^-$ the number of negative eigenvalues.

Since we are only allowed integer change of bases, the problem is
more difficult than this.  We still do have $b_2^+$ and $b_2^-$, but
several matrices may have the same values for $b_2^+$ and $b_2^-$ but
not be equivalent.

In particular, the classification of definite symmetric bilinear forms
($b_2^+=0$ or $b_2^-=0$) is not known at all, but it is known that the
number of these forms of even moderate size is quite large.  If the
classification of simply connected four-dimensional manifolds depends
on understanding this classification, we are in some
trouble.\footnote{Not as much trouble as we were with $\pi_1(X)$.  This
classification is not algorithmically impossible; just poorly
understood.}

The classification of the indefinite case (neither $b_2^+$ nor $b_2^-$
is zero) is actually much better: we know the classification of these
completely.  When at least one diagonal element is odd, it is possible
to change the basis so that the matrix is diagonal with only $1$'s and
$-1$'s on the diagonal; and when all diagonal elements are even, the
basis can be chosen so that the matrix breaks up into $2\times 2$
blocks and $8\times 8$ blocks, where the $2\times 2$ blocks are the
matrix
\[H=\left(\begin{array}{cc}
0&1\\
1&0
\end{array}\right)\]
and the $8\times 8$ blocks are each the Cartan matrix for the Lie group $E_8$:
\[E_8=\left(\begin{array}{cccccccc}
2&-1&0&0&0&0&0&0\\
-1&2&-1&0&0&0&0&0\\
0&-1&2&-1&0&0&0&0\\
0&0&-1&2&-1&0&0&0\\
0&0&0&-1&2&-1&0&-1\\
0&0&0&0&-1&2&-1&0\\
0&0&0&0&0&-1&2&0\\
0&0&0&0&-1&0&0&2
\end{array}\right)\]
The matrix $E_8$ is definite, so there must be at least one $H$, or we
would be considering the definite case above, instead of the indefinite
case.

\begin{figure}
\begin{center}
\begin{tabular}{c|c|c|}
    &Indefinite& Definite\\\hline
Odd &
\parbox{2in}{\begin{center}
$m(1)\oplus n(-1)$\\
$m, n\ge 1$
\end{center}}
&
\parbox{2in}{$\pm I$, $E_8\oplus (1)$, many more (unknown)}
\\\hline
Even&
\parbox{2in}{\begin{center}
$mH\oplus nE_8$\\
$m\ge 1$, $n\ge 0$
\end{center}}
&\parbox{2in}{\begin{center}
$nE_8$, $SO(32)$, Leech lattice, many more (unknown)
\end{center}}\\\hline
\end{tabular}
\end{center}
\caption{Classification of symmetric, integer-valued bilinear forms of
determinant $\pm 1$, up to change of basis.  ``Even'' means all the diagonal
elements are even, and ``odd'' means at least one diagonal element is odd.}
\label{fig:forms}
\end{figure}

Note that $b_2=b_2^++b_2^-$.  As usual, $\sigma(X^4)=b_2^+ - b_2^-$ is
called the {\em signature.}  If the orientation of the manifold is reversed,
the matrix is replaced by its negative, and therefore the $b_2^+$ and
$b_2^-$ reverse roles.  So $b_2^+$ and $b_2^-$ are not really topological
invariants, but $|\sigma|$ is.  Alternately, we can try to classify
manifolds together with their orientations, and then we have $b_2^+$ and
$b_2^-$ as invariants of manifolds with orientation.

\begin{figure}
\begin{center}
\begin{tabular}{ccccc}
Manifold&$b_2$&$I$&$b_2^+$&$b_2^-$\\\hline
$S^4$&$0$&$()$&$0$&$0$\\
$S^2\times S^2$&$2$&$H$&1&1\\
$\cc P^2$&1&$(1)$&1&0\\
$\overline{\cc P^2}$&1&$(-1)$&0&1\\
K3&22&$3H\oplus 2E_8$&19&3\\
$(\cc P^2)^{\#m}\#(\overline{\cc P^2})^{\#n}$
   &$m+n$&$m(1)+n(-1)$&$m$&$n$\\
$\mbox{K3}^{\#m}\#(S^2\times S^2)^{\#n}$&
$22m+2n$&$(3m+n)H\oplus (2m)E_8$&$19m+n$&$3m+n$\\
\end{tabular}
\end{center}
\caption{Some intersection forms of simply connected four-dimensional
manifolds}
\label{fig:intersectionexamples}
\end{figure}

Now for some examples of intersection forms.  The four-sphere $S^4$
has $b_2=0$, so the matrix is the empty zero-by-zero matrix.  For
$S^2\times S^2$, $b_2=2$, and the intersection form is $H$
above, and $b_2^+=1$, and $b_2^-=1$, so $\sigma=0$.  For $\cc P^2$, we
have $b_2=1$, and the intersection form is $(1)$ (assuming the usual
orientation on $\cc P^2$).  Then $b_2^+=1$ and $b_2^-=0$, and
$\sigma=1$.  We call $\overline{\cc P^2}$ the manifold $\cc P^2$ with
the reverse orientation, so that $b_2^+=0$, $b_2^-=1$, and
$\sigma=-1$.

We can also take the connected sum of two simply connected
four-dimensional manifolds, resulting in a new simply connected
four-dimensional manifold.  The second Betti number of the resulting
manifold is $b_2(X\#Y)=b_2(X)+b_2(Y)$, and similarly $b_2^+$, $b_2^-$,
and $\sigma$ are additive.  Furthermore, the resulting intersection
form can be put into blocks
\[\left(\begin{array}{cc}
I_X &0\\
0&I_Y
\end{array}\right)\]
where $I_X$ is the matrix for the intersection form on $X$ and $I_Y$ is the
matrix for the intersection form on $Y$.  So if we take connected sums of
copies of $\cc P^2$ and $\overline{\cc P^2}$, we can form a manifold with
intersection form with arbitrarily many $1$'s and $-1$'s down the diagonal,
with the rest of the matrix zero.  If we only use $\cc P^2$'s, we get the
identity matrix, and if we only use $\overline{\cc P^2}$'s, we get minus the
identity matrix.

Note that $S^2\times S^2$ and $\cc P^2\#\overline{\cc P^2}$ have the
same homology groups ($b_2=2$ in both cases) but their intersection
forms are different.

A remarkable manifold is the K3 surface (the four-dimensional equivalent
to Calabi--Yau manifolds).  The K3 surface can be described as the solutions to
\[x^4+y^4+z^4=1\]
where $x$, $y$, and $z$ are complex numbers, and where we compactify
the parts that go to infinity by considering $(x,y,z)$ as coordinates
on $\cc P^3$.  The K3 surface is a smooth four-dimensional manifold,
simply connected, and has $b_2=22$, with $b_2^+=19$, $b_2^-=3$, and
$\sigma=16$.  The intersection form is a $22\times 22$ matrix, which
can be put into block diagonal form with three $H$'s and two $E_8$
blocks.

Because of the block structure, we might suspect the K3 surface is a
connected sum of various pieces, two that have $E_8$ as their
intersection form, and three that have $H$ as their intersection form, and
in fact this might have been supposed before the 1980s\footnote{Actually,
because of Rokhlin's theorem, the two $E_8$'s cannot be separated, but
we might have thought that was the only restriction.}, but
this turns out not to be true, as we now know from what is described below.

There was a little bit known beyond this before the 1980s, but not much.
Basically nothing was known about which intersection forms were possible,
and basically nothing was known about whether it was possible for two
manifolds to have the same intersection form (implicit is that they would
have the same $b_2$, and in particular, the same homology and cohomology).
\footnote{There was a little more than what I have mentioned here:
a theorem by Rokhlin, and an invariant by Kirby and Siebenmann, and so on.
These are summarized in Kirby's book mentioned above.\cite{Ki}}

The reader may have noticed that we have spent some time with homology
and cohomology groups, and a lot of time with the intersection form,
but we have not discussed the higher homotopy groups.  The higher
homotopy groups are in general too difficult to calculate, but in the
end turn out not to give new information anyway for simply connected
four-dimensional manifolds.

For a more detailed account of this section, see Kirby's book {\em The
Topology of 4-Manifolds}\cite{Ki}.

\section{The two breakthroughs in the 1980s}
There were two breakthroughs in the 1980s that suddenly added
remarkable clarity to what was going on for
simply connected\footnote{The condition that the manifold be
simply connected can be somewhat loosened, and the story is fairly
similar.  For the rest of this article we assume the manifold is
simply connected, to make the notation clearer, and to avoid having
lots of complicated restrictions on the statements of the results.
Even though classification of four-dimensional manifolds is
impossible, the techniques we describe are still useful in general.}
four-dimensional manifolds, and they happened at roughly the same
time.  On the one hand was the work of Michael Freedman that was
completely topological, and on the other was the work of Simon
Donaldson that used instantons.  These two breakthroughs were
complementary in the sense that they addressed two disjoint sides of
the question.

Freedman's work classified topological manifolds (where the coordinate
charts need not patch together smoothly) up to homeomorphism (for two
topological manifolds to be homeomorphic, all that is necessary is the
existence of a continuous map from one to the other with a continuous
inverse) as opposed to Donaldson's work which described what happens
to smooth manifolds (where the coordinate charts patch together
differentiably) up to diffeomorphism (so that the map relating the two
and its inverse must be differentiable).  It turns out the stories for
the smooth classification and for the non-smooth classification are
very different.  \footnote{The non-smooth classification is sometimes
referred to as the {\em topological} classification, since notions of
continuity are required but not notions of differentiability.  It is
nevertheless common for mathematicians to use the word ``topological''
in the context of the smooth classification, in phrases like
``topological invariant'', when there is no chance for confusion, and
since beyond this section the primary consideration is with the smooth
classification problem, we will sometimes use the word ``topological''
in this way.}

Freedman's work\cite{Fr}, published in 1982, showed that for simply
connected compact four-dimensional manifolds without boundary, all
intersection forms depicted in Figure~\ref{fig:forms} are possible,
and with an additional $\zz_2$-valued invariant known as the
Kirby--Siebenmann invariant, these data completely determine the
manifold up to homeomorphism.  Thus, the question of classifying
simply connected compact topological four-dimensional manifolds without boundary
up to homeomorphism was finally solved.\footnote{This does not include
the fact that intersection forms are not classified, of course.}  The
idea behind Freedman's work is to show that a more sophisticated
version of what works for dimensions five and higher actually works
for dimension four.  In dimensions five and higher, it is often
necessary to ``simplify'' a description of a manifold by finding a
complicated subset and showing it is really a ball.  The same idea
works in dimension four, except that sometimes the necessary subset is
infinitely complicated, and Freedman was able to show that such a subset
is homeomorphic (though perhaps not diffeomorphic) to a ball.  Since
this did not involve physics, we will not discuss this work further,
but a good place to learn this is in the book by Freedman and
Quinn\cite{FQ}.

On the other hand, Donaldson's work, starting with the seminal
publication\cite{D1} in 1983, dealing with smooth manifolds up to
diffeomorphism, did not result in as complete an answer, but what
Donaldson discovered resulted in a simplification along a completely
different direction.  By considering a Yang--Mills $SU(2)$ gauge field
on the four-dimensional manifold, and studying instantons, Donaldson
was able to prove that the intersection form must be either indefinite
(in which case we know how to classify such intersection forms) or
plus or minus the identity.  In other words, the situation where we
didn't know how to classify intersection forms, the case where it was
definite, is the situation where we this classification is
unnecessary, since smooth manifolds can't have them as intersection
forms anyway, with the exception of the identity and minus the
identity (see Figure~\ref{fig:donaldsonforms} and compare to the earlier
Figure~\ref{fig:forms}).

\begin{figure}
\begin{center}
\begin{tabular}{c|c|c|}
    &Indefinite& Definite\\\hline
Odd &
\parbox{2in}{\begin{center}
$m(1)\oplus n(-1)$\\
$m,n\ge 1$
\end{center}}
&
\parbox{2in}{\begin{center}
$\pm I$
\end{center}}\\\hline
Even&
\parbox{2in}{\begin{center}
$mH\oplus nE_8$\\
$m,n$ ?\\
see Figure~\ref{fig:geography}
\end{center}}
&
\parbox{2in}{\begin{center}
nothing
\end{center}}\\\hline
\end{tabular}
\end{center}
\caption{Classification of intersection forms of smooth manifolds, due
to Donaldson.  Note that the unknown areas of Figure~\ref{fig:forms} have
disappeared, and almost all entries are represented by manifolds given
in Figure~\ref{fig:intersectionexamples}.  The only situation unknown is
the number of $m$ and $n$ possible for even intersection forms.}
\label{fig:donaldsonforms}
\end{figure}

This is known as Donaldson's Theorem A, since there are other
important theorems in that paper\cite{D1}, all deriving from analyzing
the equations for instantons.  The original papers are Ref.~\cite{D1}
and Ref.~\cite{D2}.  A friendly introduction to the subject is Freed
and Uhlenbeck's book\cite{FU}, and a detailed textbook is a book by
Donaldson and Kronheimer\cite{DK}.

Before giving a sense for how Donaldson's Theorem A was proved, and before
giving other important uses of this technique, let us recall a few
things about instantons.

\section{Instantons}
Consider a pure $SU(2)$ gauge field theory on flat $\re^4$, as
described in standard textbooks like Peskin and Schroeder\cite{PS}.
Let $i\sigma^a$ be the standard Pauli basis for the Lie algebra of
$SU(2)$, where $a=1, \dots, 3$.  Let $A_\mu=A^a_\mu \sigma_a$ be an
$SU(2)$ connection, with $\mu = 1,\dots, 4$ a spatial index, and
$F^a_{\mu\nu}=\partial_\mu A^a_\nu - \partial_\nu A^a_\mu +
\epsilon^{abc} A^b_{\mu} A^c_{\nu}$ is its curvature tensor, so that
\[F_{\mu\nu}=A_{[\mu,\nu]}+[A_\mu, A_\nu].\]
Consider the action
\[S=\int_{\re^4}\|F\|^2\,d^4x=\int_{\re^4} F_{\mu\nu}^aF^{\mu\nu}_a\,d^4x.\]
If we replace the Lorentzian $(-+++)$ metric with the Euclidean
$(++++)$ metric, we can obtain classical minima of the action above.
These are called {\em instantons}, and are useful in calculating
tunneling amplitudes\cite{Cole} (the rotation from time to imaginary
time is what is involved in the WKB approximation).

We care not about $\re^4$ but about arbitrary compact manifolds
($\re^4$ is not compact).  The question of
finding instantons is basically unchanged, except when $\re^4$ is
replaced by a non-trivial manifold, we need to consider some
topological considerations.  Namely, the gauge field corresponds to a
vector bundle $E$ (in this case, a two dimensional complex vector bundle)
on the manifold.  The connection is locally defined on coordinate
patches, and transforms as we go from one patch to another by gauge
transformations.

For each such vector bundle $E$ over our manifold $X^4$ we can associate the
second Chern class
\[c_2(E)=-\frac{1}{8\pi^2}\int_{X^4} F_{\mu\nu}^a \tilde{F}^{\mu\nu}_a\]
which is an integer.\footnote{More precisely, $c_2(E)$ is the
four-form in the integrand; it is an element of $H^4(X^4)\cong \zz$.
The isomorphism $H^4(X^4)\cong \zz$ is realized by taking the
integral.}  The second Chern class is defined above in terms of the
connection $A$, through its curvature $F$, but in fact it is
independent of the connection and only depends on the vector bundle $E$.  The
first Chern class $c_1(E)$, incidentally, is zero because the group is
$SU(2)$.  In the $U(1)$ gauge theory, the first Chern class in
generally non-zero and measures the monopole charge for a Dirac
monopole.  There are higher Chern classes but they are all zero for
$SU(2)$.

It turns out that the second Chern class completely classifies the
vector bundle topologically, so that there is a unique vector bundle
up to topological vector bundle isomorphism for every integer value of
$c_2$.  The trivial bundle has $c_2=0$.

For each vector bundle $E$, we can look for connections $A$ that minimize the
Yang--Mills action.  One choice might be the trivial connection
$A=0$, which gives rise to the action being equal to zero.  This is
clearly an absolute minimum, because the action in our case cannot
be negative.  But this trivial connection only exists in the trivial
bundle.  When we plug in this connection into the formula for the
second Chern class, we get $c_2=0$.  More generally, any flat connection is
a minimum, but also exists only in the trivial bundle with $c_2=0$.

For other vector bundles, the minima are not as obvious.  The trick to
understanding these minima is to split the curvature $F$ into the $+1$
and $-1$ eigenvalues of the duality operator $*$, where
$*F=\tilde{F}$.  We define $F^+=\frac{1}{2}(F+\tilde{F})$ and
$F^-=\frac{1}{2}(F-\tilde{F})$.  Then $F=F^+ + F^-$, where $*F^+=F^+$
and $*F^-=-F^-$.  Furthermore, $F^+$ and $F^-$ are orthogonal.  The
formula for $c_2$ gives
\begin{eqnarray*}
c_2(E)&=&-\frac{1}{8\pi^2} \int_{X^4} (F^+ + F^-)_{\mu\nu} (*(F^+ + F^-))^{\mu\nu}\\
&=&-\frac{1}{8\pi^2} \int_{X^4} F^+_{\mu\nu} (*F^+)^{\mu\nu} + F^-_{\mu\nu}
(*F^-)^{\mu\nu}\\
&=&-\frac{1}{8\pi^2} \int_{X^4} F^+_{\mu\nu}F^{+\mu\nu} - F^-_{\mu\nu}F^{-\mu\nu}\\
&=&\frac{1}{8\pi^2} \int_{X^4} -\|F^+\|^2 + \|F^-\|^2
\end{eqnarray*}
while the formula for the action is
\begin{eqnarray*}
S&=\int_{X^4} (F^++F^-)_{\mu\nu}(F^++F^-)^{\mu\nu}\\
&=\int_{X^4} \|F^+\|^2+\|F^-\|^2.
\end{eqnarray*}

Thus we see that when $c_2(E)<0$, the action is minimized when
$F^-=0$, so that for instantons, $*F=F$ (in which case we call $F$
{\em self-dual}) and when $c_2(E)>0$, the action is minimized when
$F^+=0$, so that instantons have $*F=-F$ (in which case we call $F$
{\em anti-self-dual}, and we sometimes call such solutions {\em
anti-instantons}).  When $c_2(E)=0$, the action is minimized when
$F=0$, which we observed before.

Suppose we have an instanton with $c_2(E)=1$.  We view this as a
minimum of the action.  When we ask the question as to why this is the
minimum when the connection $A=0$ clearly gives a lower value for the
action, the answer is that $A=0$ does not exist in our bundle.  To
``decay'' from our instanton to zero would require that we ``tear''
our bundle first to untwist it.  This is what we mean when we say that
the instanton cannot decay for topological reasons.  The number
$c_2(E)$ (more conventionally, $-c_2(E)$) is called the {\em instanton
number} of the solution, and we imagine that instantons with
$c_2(E)=2$ are in some sense ``non-linear'' combinations of two
instantons with $c_2(E)=1$.  When we combine a solution with $c_2=-1$
(an instanton) with a solution with $c_2=1$ (an anti-instanton), they
can cancel and flow down to a flat connection.

Before I continue, let us consider a point about these solutions.  The
critical points of the action could have been found using very
standard classical techniques using the calculus of variations, and it
turns out that this gives us the equation $D(*F)=0$.  Since the
connection $A$ is the dynamical field we are interested in, and $F$
involves a derivative of $A$, we see that $D(*F)=0$ is a
second-order equation.  Instead, we have just derived the equations
$*F=F$ or $*F=-F$ which are first-order equations in $A$.
The difference is that these first-order equations hold for
only the absolute minima, and do not hold for other relative (local)
minima, nor do they hold for any sort of ``saddle'' points of the
action.  The Bianchi identities $DF=0$ can assure us that any
solution to $*F=\pm F$ will also satisfy $D(*F)=0$, but the
reverse is not true.  It turns out that there are many critical points
(those that satisfy the Euler--Lagrange equation $D(*F)=0$) that
are not absolute minima (instantons, satisfying $*F=\pm F$).  Now
the physics reasons for studying instantons really does prefer
absolute minima, anyway, so it could be argued that $*F=\pm F$ is
really what we want to solve.  From the mathematical perspective, we
get to choose whatever equations we happen to like, and it is
solutions to $*F=\pm F$, not $D(*F)=0$, that led to new
developments in four-dimensional topology, so that is what we will
focus on here.

We now consider instantons on $S^4$.  Readers who are familiar with
instantons on $\re^4$ will see many similarities.  The reason is that
the Yang--Mills action above has a conformal symmetry, and there is a
conformal map from $\re^4$ to $S^4$ that covers everything except for
one point.  The work of Atiyah, Drinfeld, Hitchin, and Manin\cite{ADHM}
gives an explicit description of these instantons, and we will here
describe their results for $c_2(E)=-1$.

In the case $c_2(E)=-1$, we are looking for self-dual connections on
$E$, which involves solving the differential equation $*F=F$ for $A$.
It turns out that the set of instantons on a bundle with $c_2(E)=-1$
on $S^4$, modulo gauge symmetry, is naturally a five-dimensional
non-compact manifold.  More specifically, it is a five-dimensional
open ball.  We call this set the {\em moduli space}.  It turns out
we can identify $S^4$ with the missing boundary of the ball in a sense
I will describe in a moment.

But before doing that, we should first consider how it came to be that
the set of minima is not unique.  Usually, a function has a unique absolute
minimum.  It is possible to have functions that have many absolute minima,
by arranging it so that many points take on the same minimum value of the
function.  But we usually regard this as an unusual phenomenon, and in
the world of physics, where the formulas are given to us by nature rather
than specifically dreamed up to have multiple minima, we should expect
there to be only one absolute minimum.  If we see more than one absolute
minimum, this is a phenomenon to be explained.

There are, indeed, circumstances in physics that give multiple absolute
minima, and even continuous families of absolute minima, but these are
usually explained by the existence of a group of symmetries.  Take,
for example, the Higgs mechanism in a $\phi^4$ theory.  The theory has
a spherical symmetry, and so the set of minima might be a sphere, and
small perturbations that preserve this symmetry will still have a
spherical set of minima.

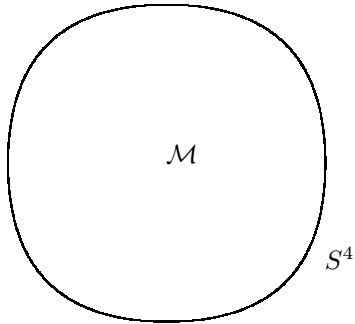
\begin{figure}
\begin{center}
\begin{picture}(300,140)
\qbezier(90,70)(90,130)(150,130)
\qbezier(150,130)(210,130)(210,70)
\qbezier(210,70)(210,10)(150,10)
\qbezier(150,10)(90,10)(90,70)
\put(150,70){{$\moduli$}}
\put(210,30){{$S^4$}}
\end{picture}
\end{center}
\caption{Schematically drawn here, the moduli space of instantons
with $c_2(E)=-1$ on $S^4$ is a five-dimensional ball, with the original
$S^4$ as its boundary.}
\end{figure}

In the case of instantons on $S^4$, the existence of many minima
can also be explained by symmetry.  There is the gauge symmetry, but
recall that we have already quotiented out by this symmetry.  But
there are also conformal symmetries of $S^4$, and since the action is
conformally invariant, these conformal symmetries will take instantons
to other instantons.  In fact, the conformal symmetries of $S^4$ are
enough to explain the entire set of solutions in this case.  Therefore,
from one solution, we can use the conformal symmetries to explain the
entire moduli space.  The fact that there are no other solutions
for $c_2(E)=-1$ was shown by Atiyah and Ward\cite{AW}.

Taking this idea of using the conformal symmetry, we can take a
conformal symmetry that flows all of $S^4$ concentrating more and more
of it closer to any given point on $S^4$.  The effect of this is to
concentrate the instanton near a given point of $S^4$.  This explains why the
boundary of the set of solutions is $S^4$.  The conformal symmetry
that concentrates most of $S^4$ near a point $p\in S^4$ will also move
instantons in the moduli space (recall it is a five-dimensional ball)
near a corresponding point on its boundary.  The limiting connection
is degenerate, and in a sense that is reminiscent of a Dirac delta
function, is flat everywhere on $S^4$ except at $p$, where it has
infinite curvature.  Thus we can add to our moduli space these extra
limiting connections, thereby turning our non-compact ball to a compact ball
with boundary.\footnote{These limiting degenerate configurations are
sometimes called {\em small instantons}, and while physicists are used to
viewing them as instantons of a special kind, mathematicians tend not to
view them as instantons, since $A_\mu$ is not even well-defined at the
point $p$.  But it is possible to define a ``small instanton'' and add these
small instantons to the moduli space in a natural way.  The result makes the
moduli space compact and this process is called {\em compactifying the
moduli space}.}

More generally, the moduli space of instantons on a
four-dimensional manifold $X^4$, with $c_2(E)<0$ given, is a
manifold\footnote{When we say, ``manifold'', we perhaps should say
``orbifold'' instead, since singularities due to quotients of group
actions sometimes occur in moduli spaces, as we will see later.  But
referring to the moduli space as a manifold is entrenched in the
literature and we imagine that it is a manifold, but perhaps with a
few singularities.  Besides this, we already saw in the case of $S^4$
that the moduli space is not necessarily compact, until we include the
small instantons which may add a boundary.} of dimension
\begin{equation}
d=-8c_2(E)-3(1-b_1(X^4)+b_2^-(X^4)).
\label{eqn:dim1}
\end{equation}
This formula is obtained by the Atiyah--Singer index theorem, by
viewing the self-dual equations as zeros of a differential operator,
together with a suitable gauge-fixing condition like $d^*(A-A_0)=0$
once a fixed reference connection $A_0$ is identified.

Similarly, when $c_2(E)>0$, we are solving the anti-self-dual equation
$*F=-F$, with the same gauge-fixing condition, and the Atiyah--Singer
index theorem gives the dimension as
\begin{equation}
d=8c_2(E)-3(1-b_1(X^4)+b_2^+(X^4)).
\label{eqn:dim2}
\end{equation}

The dimension may be zero, in which case the moduli space would be a set of
points, or the dimension may be negative, in which case the moduli space
will be empty (so that there would be generically no instantons with that
value of $c_2$).

If this dimension $d$ is positive, we should in general have many
absolute minima, and we might want to explain why this is the case.  We no
longer have the conformal symmetry of $S^4$ in general, so we have no
reason to suspect multiple solutions.  In fact, with many known cases,
we see no obvious symmetry in the moduli space.  This is an example
of a situation where the dictum that multiple minima must come from a
group is unfounded.  The reason a function $f:\re^n \to \re$ typically
has a unique absolute minimum, or at least zero-dimensional relative minima,
is that the criterion $\nabla f=0$ gives $n$ equations.  A system of
$n$ equations and $n$ variables typically results in a zero-dimensional
set of solutions.  If instead of $\nabla f=0$ we had $n-1$ equations, we
would expect a one-dimensional set of solutions.

In our case, the action $S$ has an infinite dimensional domain
(the set of connections), and the criterion of $\nabla S=0$ results in
infintely many equations.  But we can no longer make any sense of
comparing the number of equations and number of variables.  In fact,
in general, infinite dimensional problems like this have many
pathologies, and there is nothing more we can say.  But in our
particular case, because the differential equations are elliptic, we
have many nice results.  For instance, even though we have an infinite
number of ``variables'' and an infinite number of ``equations'', we
have a well-defined notion of the difference between the two
dimensions, and this number is called the index.  This is the dimension
of the set of solutions, and is what the Atiyah--Singer index theorem
calculates.  This is the number given above for the dimension of
the moduli space.

In other words, for infinite-dimensional problems like this, we must
discard our intuition that was based on finite-dimensional problems,
and if we have an elliptic differential equation (as in our case) we
can be thankful that the intuition need not be completely discarded,
but only modified.  An example that may be more familiar to you is the
fact that the minima of
\[\int_{X^4} \|d\omega\|^2+\|d^*\omega\|^2\]
where $\omega$ ranges over $k$-forms are those $k$-forms $\omega$ that
are harmonic, that is, satisfy $\Delta \omega=0$.  The set of these is
the $k$-th cohomology group $H^k(X^4;\re)$, and this is a vector space
of dimension $b_k$.\footnote{Some may point out that this has a
symmetry, too, in that harmonic forms act on the set of solutions by
addition.  But this begs the point, since if there were not an
infinite family of harmonic forms, there would be no group to act on
the set.  Anyway, the point is that analogous situations come up in
more elementary settings, and the Atiyah--Singer index theorem
predicts the correct dimension of the space.}

There is no group that guarantees a non-zero-dimensional family of
solutions: the ``correct'' dimension of the set of minima is simply
given by the Atiyah--Singer index theorem.  If you plug in $S^4$ and
$c_2(E)=-1$ in the dimension formula (\ref{eqn:dim1}), (note that for
$S^4$ we have $b_1=0$ and $b_2^-=0$) we get $d=5$, which says that the
five-dimensionality of the moduli space there is not really a
consequence of the conformal symmetry group after all, in the sense
that the moduli space would continue to be five-dimensional even if we
were to slightly perturb the metric on $S^4$ so that it no longer has
conformal symmetry.

For a more detailed description of the moduli space of instantons on
$S^4$, see Ref.~\cite{FU} and Ref.~\cite{DK}.

\section{Donaldson's Theorem A}
As mentioned above, Donaldson's Theorem A states:

\begin{theorem}
Let $X^4$ be a simply connected compact four-dimensional manifold (no
boundary) with definite intersection form.  Then its intersection
form, in some basis, is plus or minus the identity matrix.
\end{theorem}

A rough proof goes as follows: By changing the orientation on $X^4$ we
can assume that the intersection form is positive-definite.  Then
$b_2^-=0$.  For simply connected manifolds, we saw above that $b_1=0$.  Then
if we are interested in the bundle $E$ over $X^4$ with $c_2(E)=-1$, we
see that the formula for the dimension of the moduli space
(\ref{eqn:dim1}) gives us that the moduli space of instantons will be
a five-dimensional manifold.
\begin{figure}
\begin{center}
\begin{picture}(300,200)
\qbezier(30,100)(30,155)(50,155)
\qbezier(50,155)(80,155)(80,100)
\qbezier(80,100)(80,45)(50,45)
\qbezier(50,45)(30,45)(30,100)
\put(20,35){{$X^4$}}
\qbezier(50,155)(90,155)(100,165)
\qbezier(100,165)(110,175)(190,175)
\qbezier(190,175)(210,175)(220,165)
\qbezier(220,165)(230,155)(250,155)
\qbezier(250,155)(230,155)(230,125)
\qbezier(230,125)(230,95)(250,95)
\qbezier(250,95)(230,95)(230,65)
\qbezier(230,65)(230,35)(250,35)
\qbezier(250,35)(230,35)(210,15)
\qbezier(210,15)(200,5)(100,5)
\qbezier(100,5)(80,5)(80,25)
\qbezier(80,25)(80,45)(50,45)
\put(150,100){{$\moduli$}}
\put(270,90){{singularities}}
\put(265,95){\vector(-1,0){10}}
\put(265,115){\vector(-1,4){10}}
\put(265,75){\vector(-1,-4){10}}
\end{picture}
\end{center}
\caption{The moduli space of instantons with $c_2(E)=-1$ on $X^4$ with
$b_2^-=0$.  Note the boundary as $X^4$ itself, and the $m/2$
singularities.  This drawing is intended to be schematic or
suggestive: the four-dimensional manifold $X^4$ is drawn as a circle,
and the five-dimensional moduli space is drawn as a surface.}
\end{figure}
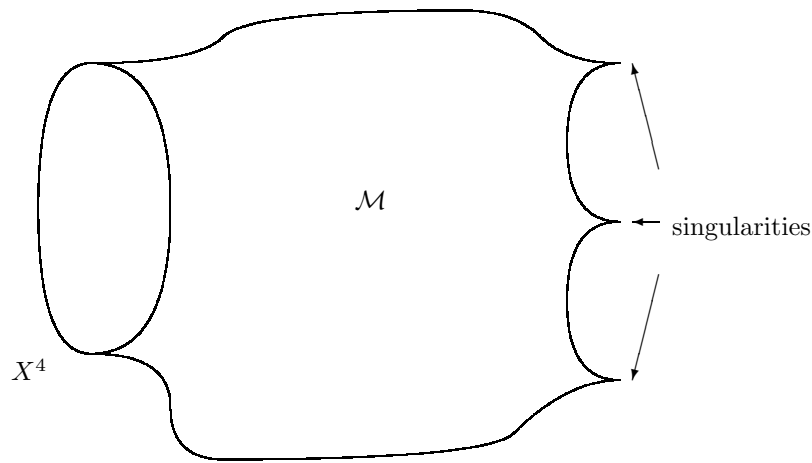

Analogously to the case for $S^4$, where $S^4$ could be viewed as the
boundary of the moduli space, we can similarly ``compactify'' the
moduli space by including small instantons (the set of small instantons
looks like a copy of $X^4$) so that the resulting
moduli space is a five-dimensional manifold with boundary $X^4$.  The
procedure is analogous to the one mentioned above, and was developed
by Karen Uhlenbeck\cite{U}.

Now I mentioned under my breath that the moduli space may not be quite
a manifold, because it may have singularities.  It turns out that in
the situation we are describing, there are finitely many
singularities, each isolated and locally isomorphic to a cone on
$\overline{\cc P^2}$.  They can be counted in the following way: let
$m$ be the number of elements $v\in H_2(X^4)$ so that $v^T I v=1$,
where $I$ is the intersection form of $X^4$.  Then there will be $m/2$
singularities.

These singularities come about from the fact that the gauge group does
not always act freely.  When the complex two-dimensional bundle $E$
can be split into two one-dimensional bundles $E=L_1\oplus L_2$, in
such a way that the connection $A$ turns out to be the product of
connections on each of the one-dimensional bundle factors, so that the
connection is actually a product of $U(1)$ connections, then a part of
the gauge group will fix $A$.  In particular, a constant $U(1)$ gauge
transformation will leave this reducible connection $A$ invariant.
Such connections are called {\em reducible}, and if this does not
occur, we call it {\em irreducible}.

The result is that when $A$ is reducible, and we quotient by the
global gauge group, there will be the kind of singularity mentioned
above: a cone on $\overline{\cc P^2}$.  Studying
the self-dual equations for connections of this special type shows
that each splitting of $E$ into two factors contributes a unique
reducible connection, and for this splitting to happen
$c_1(L_1)+c_1(L_2)=0$ and $c_1(L_1)^T I c_1(L_2) =c_2(E)$.  So these
correspond to elements $v=c_1(L_1)\in H_2(X^4)$ so that $v^T I v=1$,
and this is a one-to-one correspondence up to swapping the roles of
$L_1$ and $L_2$.  This explains the number of singularities.

These singularities are isolated and do not occur on the glued-in
$X^4$.

Therefore, we can take our moduli space of instantons and modify it as
follows: first, glue in the $X^4$ so that the moduli space becomes a compact
manifold with boundary and with singularities.  Then excise a small
open ball around each of the $m/2$ singularities.  What we now have is
a five-dimensional manifold with $X^4$ as one boundary component, and
$m/2$ other boundary components, each of which is a $\overline{\cc P^2}$.

Now, an old topological theorem is that when a union of
four-dimensional manifolds is the boundary of a five-dimensional
manifold, the sum of their signatures is zero.\footnote{For this theorem
to apply we need the five-dimensional manifold to be orientable.  It turns
out that these moduli spaces are always orientable.}  In our case, this
means $\sigma(X^4)+\frac{m}{2}\sigma(\overline{\cc P^2})=0$.  Since
$\sigma(\overline{\cc P^2})=-1$, this means $\sigma(X^4)=m/2$.  Since
the intersection form for $X^4$ is positive definite, $\sigma=b_2$,
so there are $b_2$ singularities in the original moduli space.  Therefore,
there are $b_2$ solutions to $v^TIv=1$.

A few words about the $v\in H_2(X^4)$ with  $v^T I v=1$.  Note
that if $v$ satisfies this, so does $-v$, which partly explains the
naturality of dividing by 2.  Furthermore, if $v$ and $w$ are two
such, and $v\not=\pm w$, it turns out that $v^T I w=0$.

Choose one $v$ from each $\pm$ pair of solutions to $v^TIv=1$.
The collection of these $v$ will then form a basis, and in this basis,
$I$ will be the identity matrix.

This proves Donaldson's Theorem A.

\section{Other questions about existence}
The question of classification of smooth manifolds might be split into
two questions: For each intersection form, do there exist manifolds
with that intersection form?  And for each intersection form, how many
manifolds have that same intersection form?  The first question might
be viewed as an ``existence'' question, and the second question might
be viewed as a ``uniqueness'' question.

What Donaldson's Theorem A does is eliminate a great many intersection
forms from consideration, showing that manifolds do not exist that
have those intersection forms.  So it weighs in on the ``existence''
question.  The extent to which it works is apparent when you
realize that the only intersection forms left to consider are
blockwise combinations of $H$'s and $E_8$'s, and diagonal matrices
with $\pm 1$ on the diagonal (see Figure~\ref{fig:donaldsonforms}).
\begin{figure}
\begin{picture}(345,150)
\put(0,0){\vector(1,0){345}}
\put(0,0){\vector(0,1){140}}
\put(0,0){\circle*{4}}
\put(15,0){\circle*{4}}
\put(30,0){\circle*{4}}
\put(45,0){\circle*{4}}
\put(60,0){\circle*{4}}
\put(75,0){\circle*{4}}
\put(90,0){\circle*{4}}
\put(105,0){\circle*{4}}
\put(120,0){\circle*{4}}
\put(135,0){\circle*{4}}
\put(150,0){\circle*{4}}
\put(165,0){\circle*{4}}
\put(180,0){\circle*{4}}
\put(195,0){\circle*{4}}
\put(210,0){\circle*{4}}
\put(225,0){\circle*{4}}
\put(240,0){\circle*{4}}
\put(255,0){\circle*{4}}
\put(270,0){\circle*{4}}
\put(285,0){\circle*{4}}
\put(300,0){\circle*{4}}
\put(315,0){\circle*{4}}
\put(330,0){\circle*{4}}

\put(45,20){\circle*{4}}
\put(60,20){\circle*{4}}
\put(75,20){\circle*{4}}
\put(90,20){\circle*{4}}
\put(105,20){\circle*{4}}
\put(120,20){\circle*{4}}
\put(135,20){\circle*{4}}
\put(150,20){\circle*{4}}
\put(165,20){\circle*{4}}
\put(180,20){\circle*{4}}
\put(195,20){\circle*{4}}
\put(210,20){\circle*{4}}
\put(225,20){\circle*{4}}
\put(240,20){\circle*{4}}
\put(255,20){\circle*{4}}
\put(270,20){\circle*{4}}
\put(285,20){\circle*{4}}
\put(300,20){\circle*{4}}
\put(315,20){\circle*{4}}
\put(330,20){\circle*{4}}

\put(90,40){\circle*{4}}
\put(105,40){\circle*{4}}
\put(120,40){\circle*{4}}
\put(135,40){\circle*{4}}
\put(150,40){\circle*{4}}
\put(165,40){\circle*{4}}
\put(180,40){\circle*{4}}
\put(195,40){\circle*{4}}
\put(210,40){\circle*{4}}
\put(225,40){\circle*{4}}
\put(240,40){\circle*{4}}
\put(255,40){\circle*{4}}
\put(270,40){\circle*{4}}
\put(285,40){\circle*{4}}
\put(300,40){\circle*{4}}
\put(315,40){\circle*{4}}
\put(330,40){\circle*{4}}

\put(135,60){\circle*{4}}
\put(150,60){\circle*{4}}
\put(165,60){\circle*{4}}
\put(180,60){\circle*{4}}
\put(195,60){\circle*{4}}
\put(210,60){\circle*{4}}
\put(225,60){\circle*{4}}
\put(240,60){\circle*{4}}
\put(255,60){\circle*{4}}
\put(270,60){\circle*{4}}
\put(285,60){\circle*{4}}
\put(300,60){\circle*{4}}
\put(315,60){\circle*{4}}
\put(330,60){\circle*{4}}

\put(180,80){\circle*{4}}
\put(195,80){\circle*{4}}
\put(210,80){\circle*{4}}
\put(225,80){\circle*{4}}
\put(240,80){\circle*{4}}
\put(255,80){\circle*{4}}
\put(270,80){\circle*{4}}
\put(285,80){\circle*{4}}
\put(300,80){\circle*{4}}
\put(315,80){\circle*{4}}
\put(330,80){\circle*{4}}

\put(225,100){\circle*{4}}
\put(240,100){\circle*{4}}
\put(255,100){\circle*{4}}
\put(270,100){\circle*{4}}
\put(285,100){\circle*{4}}
\put(300,100){\circle*{4}}
\put(315,100){\circle*{4}}
\put(330,100){\circle*{4}}

\put(270,120){\circle*{4}}
\put(285,120){\circle*{4}}
\put(300,120){\circle*{4}}
\put(315,120){\circle*{4}}
\put(330,120){\circle*{4}}

\put(102,56){{?}}
\put(117,56){{?}}
\put(132,76){{?}}
\put(147,76){{?}}
\put(162,76){{?}}
\put(162,96){{?}}
\put(177,96){{?}}
\put(192,96){{?}}
\put(207,96){{?}}
\put(192,116){{?}}
\put(207,116){{?}}
\put(222,116){{?}}
\put(237,116){{?}}
\put(252,116){{?}}

\put(0,15){\line(1,0){36}}
\put(36,15){\line(0,1){120}}
\put(15,75){{\parbox{2in}{\begin{sideways}{Excluded by Donaldson}\end{sideways}}}}
\put(38,30){\line(0,1){105}}
\put(38,30){\line(1,0){45}}
\put(83,30){\line(0,1){22}}
\put(83,52){\line(3,2){117}}
\put(60,100){\shortstack{Excluded by\\Seiberg--Witten}}
\put(260,-27){{Number of $H$'s}}
\put(-8,144){{Number of $E_8$'s}}
\put(-2,0){\line(1,0){4}}
\put(-8,-4){{$0$}}
\put(-2,10){\line(1,0){4}}
\put(-8,6){{$1$}}
\put(-2,20){\line(1,0){4}}
\put(-8,16){{$2$}}
\put(-2,30){\line(1,0){4}}
\put(-8,26){{$3$}}
\put(-2,40){\line(1,0){4}}
\put(-8,36){{$4$}}
\put(-2,50){\line(1,0){4}}
\put(-8,46){{$5$}}
\put(-2,60){\line(1,0){4}}
\put(-8,56){{$6$}}
\put(-2,70){\line(1,0){4}}
\put(-8,66){{$7$}}
\put(-2,80){\line(1,0){4}}
\put(-8,76){{$8$}}
\put(-2,90){\line(1,0){4}}
\put(-8,86){{$9$}}
\put(-2,100){\line(1,0){4}}
\put(-13,96){{$10$}}
\put(-2,110){\line(1,0){4}}
\put(-13,106){{$11$}}
\put(-2,120){\line(1,0){4}}
\put(-13,116){{$12$}}
\put(-2,130){\line(1,0){4}}
\put(-13,126){{$13$}}

\put(0,-2){\line(0,1){4}}
\put(-3,-12){{$0$}}
\put(15,-2){\line(0,1){4}}
\put(12,-12){{$1$}}
\put(30,-2){\line(0,1){4}}
\put(27,-12){{$2$}}
\put(45,-2){\line(0,1){4}}
\put(42,-12){{$3$}}
\put(60,-2){\line(0,1){4}}
\put(57,-12){{$4$}}
\put(75,-2){\line(0,1){4}}
\put(72,-12){{$5$}}
\put(90,-2){\line(0,1){4}}
\put(87,-12){{$6$}}
\put(105,-2){\line(0,1){4}}
\put(102,-12){{$7$}}
\put(120,-2){\line(0,1){4}}
\put(117,-12){{$8$}}
\put(135,-2){\line(0,1){4}}
\put(132,-12){{$9$}}
\put(150,-2){\line(0,1){4}}
\put(144,-12){{$10$}}
\put(165,-2){\line(0,1){4}}
\put(159,-12){{$11$}}
\put(180,-2){\line(0,1){4}}
\put(174,-12){{$12$}}
\put(195,-2){\line(0,1){4}}
\put(189,-12){{$13$}}
\put(210,-2){\line(0,1){4}}
\put(204,-12){{$14$}}
\put(225,-2){\line(0,1){4}}
\put(219,-12){{$15$}}
\put(240,-2){\line(0,1){4}}
\put(234,-12){{$16$}}
\put(255,-2){\line(0,1){4}}
\put(249,-12){{$17$}}
\put(270,-2){\line(0,1){4}}
\put(264,-12){{$18$}}
\put(285,-2){\line(0,1){4}}
\put(279,-12){{$19$}}
\put(300,-2){\line(0,1){4}}
\put(294,-12){{$18$}}
\put(315,-2){\line(0,1){4}}
\put(309,-12){{$19$}}
\put(330,-2){\line(0,1){4}}
\put(324,-12){{$20$}}
\end{picture}
\vspace*{30pt}
\caption{The possible number of $E_8$ and $H$'s.  Rokhlin's theorem
prevents us from having an odd number of $E_8$'s, and Donaldson's work
says that if we have at least two $E_8$'s, we must have at least three
$H$'s.  At the end of the review we describe Furuta's work, using
Seiberg--Witten theory, that says the number of $H$'s must be at least
the number of $E_8$'s plus one.  The work of Furuta, Kametani, and
Matsue, also using Seiberg--Witten theory, excludes $4E_8+5H$.  The
filled dots indicate which combinations are known to exist.  The
question marks indicate the current unknown areas.}
\label{fig:geography}
\end{figure}
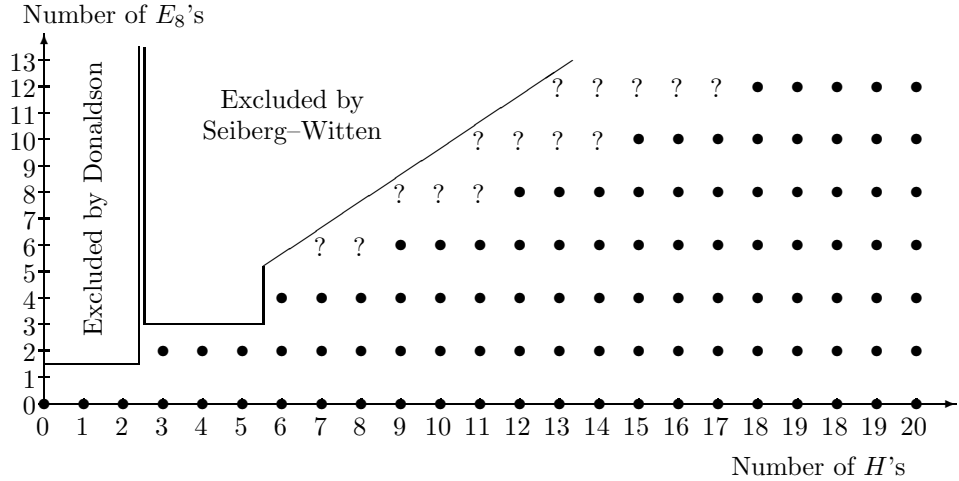

The diagonal matrices can be obtained by taking the connected sum of
copies of $\cc P^2$ and $\overline{\cc P^2}$, so these are definitely
possible.  As above, $S^2\times S^2$ has $H$, so it is possible to get
arbitrarily many $H$'s.  The K3 surface, as described above, has three
$H$'s and two $E_8$'s.  By connect summing K3 to another copy of K3,
and iterating this procedure, it is possible to get $3kH\oplus 2kE_8$.
By connect summing these by $S^2\times S^2$'s, we can get $mH\oplus n
E_8$, where $n$ is even, and $m\ge \frac{3}{2} n$.  In terms of
$\sigma$ and $b_2$, where we note that $b_2=2m+8n$ and $\sigma=8n$, we
see that we can get any $b_2$ and $\sigma$ with $b_2 \ge
\frac{11}{8}\sigma$.

The question as to whether or not the other combinations of $H$ and
$E_8$ are possible is a difficult one.  A natural conjecture, called
the {\em eleven-eighths conjecture}, is that it is impossible to have
a combination of $H$'s and $E_8$'s with $b_2 < \frac{11}{8}|\sigma|$.
This has not yet been proven nor disproven.  Donaldson's techniques
were able to make some progress (such as $m\ge 3$ when $n>0$).
Seiberg--Witten theory allowed for even more progress, described in
section 17.

\section{Uniqueness: the Donaldson invariants}
The question of discerning different manifolds that have the same
intersection form comes down to finding new invariants.  Here, too,
instantons turn out to be useful.  Using them, Donaldson defined what
are now known as {\em Donaldson invariants}, or {\em Donaldson
polynomials}.\cite{D3}  A thorough elaboration of this section can be
found in Donaldson and Kronheimer's book\cite{DK}.

To get some idea of how these might be defined, consider a
simply connected four-dimensional manifold $X^4$.  Suppose
some choice of $c_2(E)>0$ makes the dimension of the moduli space
\[d=8c_2(E)-3(1-b_1(X^4)+b_2^+(X^4))\]
equal to zero.  For instance, if $b_1=0$ (as is required for $X^4$ to
be simply connected) and $b_2^+ = 7$ (as is the case with a connected
sum of seven $\cc P^2$'s), then for the bundle $E$ over $X^4$
with $c_2(E)=3$, the moduli space of instantons would have dimension
zero, and so would be a collection of points.  These points actually
come with multiplicity and sign.  The Donaldson invariant of the
manifold $X^4$ would be the count of how many points there are in the
moduli space, counted with appropriate multiplicity and sign.

The main objection to this idea is that in order to define these
invariants, we had to assume a metric on $X^4$---and if we had used a
different metric, we would surely have different solutions to the
relevant differential equations, and hence, a different moduli space.
\begin{figure}
\begin{center}
\begin{picture}(300,150)
\put(15,15){\line(1,0){285}}
\put(15,15){\line(0,1){135}}
\put(300,15){\line(0,1){135}}
\put(7,5){{$g_0$}}
\put(292,5){{$g_1$}}
\qbezier(15,25)(100,25)(130,55)
\qbezier(130,55)(150,75)(300,75)
\qbezier(15,80)(70,80)(80,90)
\qbezier(80,90)(120,130)(300,130)
\qbezier(15,100)(60,100)(60,120)
\qbezier(60,120)(60,140)(15,140)
\put(5,22){{$+$}}
\put(305,72){{$+$}}
\put(5,77){{$+$}}
\put(305,127){{$+$}}
\put(5,97){{$-$}}
\put(5,137){{$+$}}
\end{picture}
\end{center}
\caption{A zero-dimensional moduli space under change of metric.  Note that
the Donaldson invariant would remain the same.}
\label{fig:invariant}
\end{figure}
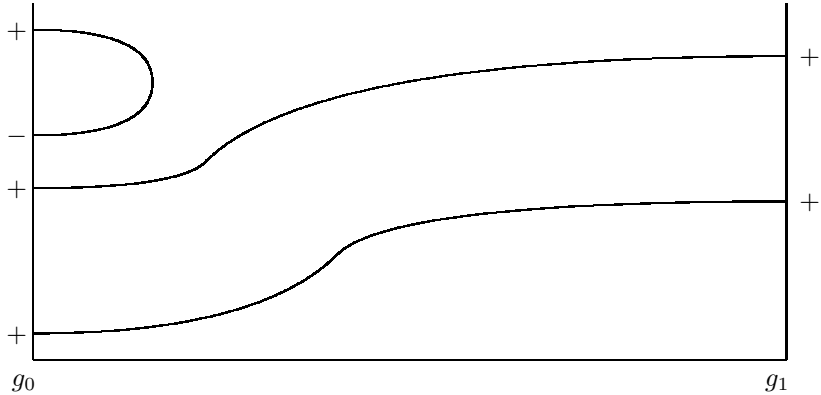

What makes this invariant a topological invariant is that this count
is independent of the metric.  The reason is that if $g_0$ and $g_1$
are two metrics on $X^4$, then since the set of metrics is connected,
we can consider a path of metrics $g_t$, $0\le t\le 1$ on $X^4$.
Then over $X^4\times [0,1]$, with the metric $g_t$ on the slice
$X^4\times\{t\}$, the moduli space over each slice joins together
to form a one-dimensional manifold.  The result is a diagram as in
Figure~\ref{fig:invariant}.

Though this looks like a Feynmann diagram, note that $t$ does not really
mean time; it is the parameter through which we are changing our metric.
Time has already been made space-like, because we are looking for instantons.
But the same kind of behavior appears: positive solutions and negative
solutions may cancel, or pairs of positive and negative solutions may appear.
So if we count these solutions with appropriate multiplicity and sign,
the number does not change.

Note, by the way, that these signs of $+$ and $-$ are {\em not} the
same thing as instantons and anti-instantons.  In our example,
$c_2(E)=3$, so each point on the moduli space is a 3-anti-instanton
solution to the differential equation, and there are no ``positive''
instantons.  The vertical axis does not represent space, either.  It
represents (schematically) the set of connections.  So a point over
$t=0$ refers to a particular 3-anti-instanton solution on $X^4$ with
metric $g_0$, and as $t$ increases, the particular solution changes
gradually.  When two points ``annihilate'', what is really going on is
two 3-anti-instanton solutions become more and more similar as the metric
is varied, and at a certain metric, they become identical, and then
the solution disappears completely.  Fans of catastrophe theory may
recognize this phenomenon.

There are some minor problems with this picture.  One problem is that
singularities may appear, and to avoid this requires that $b_2^+ > 1$.
There is nothing we can do about this except to only define the Donaldson
invariant when $b_2^+ > 1$.  This restriction thus turns up in many
results related to Donaldson theory.

The other problem is that the formula for the dimension of the moduli
space does not usually give zero, and besides, we would like to get
as many invariants in as many situations as possible.  So we should define
Donaldson invariants in the situation where the expected dimension of the
moduli space is non-zero.  This leads to the more general Donaldson invariants
which we now define.

In general, we consider the set of all irreducible connections, modulo
gauge transformation, as an infinite-dimensional manifold (call it
$\bsp=\mathcal{A}^*/\mathcal{G}$)\footnote{The $*$ superscript denotes
the fact that we are only looking at irreducible connections.}, and we
look at the moduli space of solutions modulo gauge as a
finite-dimensional submanifold (call it $\moduli_g$).  As the
metric $g$ changes, the set of solutions $\moduli_g$ moves inside the
space of connections $\bsp$.  If $g_0$ and $g_1$ are metrics on $X^4$,
then $\moduli_{g_0}$ and $\moduli_{g_1}$ are related as follows: if
$g_t$ is a path from $g_0$ to $g_1$ parameterized by $t$, then the
moduli spaces $\moduli_{g_t}$ sweep out a manifold that has as
boundary $\moduli_{g_0}$ and $\moduli_{g_1}$.  Therefore
$\moduli_{g_0}$ and $\moduli_{g_1}$ are cobordant.  The study of
submanifolds of $\bsp$ up to cobordism allows us to conclude that the
homology element defined by $\moduli_{g_0}$ in $H_d(\bsp)$ (where $d$
is the dimension of the moduli space) is the same as the homology
element defined by $\moduli_{g_1}$, or indeed, the homology element
defined by any $\moduli_g$.  Therefore the element $[\moduli]$ of
$H_d(\bsp)$ is independent of the metric and is a topological
invariant.

So this Donaldson invariant might be viewed as an element of
$H_d(\bsp)$.  It turns out that elements of $H_*(\bsp)$ can be
expressible as functions on the set of formal polynomials in the
homology of $X^4$, in a sense I will now explain.  We recall from
above that when $X^4$ is a simply connected compact four-dimensional
manifold, the homology of $X^4$ (with real coefficients) is as follows:
\begin{eqnarray*}
H_0(X;\re)&\cong&\re\\
H_1(X;\re)&\cong&0\\
H_2(X;\re)&\cong&\re^{b_2}\\
H_3(X;\re)&\cong&0\\
H_4(X;\re)&\cong&\re
\end{eqnarray*}
Let $x$ be a generator for $H_0(X^4;\re)$, and let $y_1, \dots,
y_{b_2}$ be a basis for $H_2(X^4;\re)$, and let $z$ be a generator for
$H_4(X^4;\re)$.  An argument from algebraic topology using spectral
sequences\cite{DK} shows that there is a homomorphism
$\mu:H_*(X^4;\re)\to H^{4-*}(\bsp;\re)$ so that $\mu(z)= 1\in
H^0(\bsp;\re)$, $\mu(y_1), \dots, \mu(y_{b_2})$ form a basis for
$H^2(\bsp)$, and $\mu(x)\in H^4(\bsp;\re)$.  Using wedge product we
can generate all of $H^*(\bsp;\re)$, and there are no relations other
than $\mu(z)=1$.  In other words, any element in $H^*(\bsp;\re)$ can
be written as a polynomial in $\mu(x), \mu(y_1), \dots, \mu(y_{b_2})$.
We view $\mu(x)$ as degree 4, and $\mu(y_i)$ as degree 2.  Note that
we are only working with differential forms of even degree here, so
that the wedge product is commutative.

If we have an element of the homology of $\bsp$, like $[\moduli]\in
H_d(\bsp)$, then we can pair it on any homogeneous polynomial $p$ in
$\mu(x), \mu(y_1), \dots, \mu(y_{b_2})$ of degree $d$:
\[\int_{\moduli}p(\mu(1),\mu(y_1)\wedge \dots \wedge \mu(y_{b_2})).\]
For polynomials of the wrong dimension, we define this number to be zero.
It therefore defines a function from $\re[x,y_1,\dots,y_{b_2}]$ to $\re$,
where for the sake of conciseness we can omit writing $\mu$ since
we are dealing with formal expressions anyway.

In this way, every $SU(2)$ bundle $E$ and metric $g$ on $X^4$, $\moduli_{g,E}$
gives rise to a linear function on polynomials
\[Q_{X^4,E,g}:\re[x,y_1,\dots,y_{b_2}]\to\re.\]
It does not depend on the metric $g$ as long as $b_2^+>1$.  This is
called the {\em Donaldson polynomial invariant} for $(X^4, E)$.
It is sometimes indexed not by $E$ but by the expected dimension of the
moduli space $d$, given in (\ref{eqn:dim2}), so we sometimes write
$Q_{X^4,d}$.

There are a number of technical difficulties, and a number of
restrictions due to the non-compactness of the moduli space and the
existence of singularities.  But for the most part, these problems
have been mostly solved or circumvented, as long as
$b_2^+>1$.\cite{FM,MM}

These Donaldson polynomial invariants have been calculated in a number
of circumstances, and it has been shown that many manifolds that are
indistinguishable using the ``classical'' (i.e.\ pre-1983) invariants
(such as the intersection form) turn out not to be diffeomorphic to
each other, by virtue of having different Donaldson polynomials.  From
this, and from the earlier work of Freedman mentioned above, it can be
proven that there are pairs of compact four-dimensional manifolds that are
homeomorphic to each other but not diffeomorphic to each other.

As might be expected, it is in general very difficult to calculate these
invariants, and it is not even clear if there is a general method
for calculating them, so cases where Donaldson polynomials helped distinguish
four-dimensional manifolds were rare.

In 1993, Kronheimer and Mrowka\cite{KM1} showed how to put the
various polynomials for each dimension $d$ into a generating function, called a
{\em Donaldson series}.  This is a formal non-linear function on
$H_2(X^4)$:
\[D(h)=\sum_d \frac{Q_{2d}(h^d)}{d!} + \frac{1}{2}\sum_d \frac{Q_{2d+4}(xh^d)}{d!}\]
where the sum is taken over dimensions $d$ where an $SU(2)$ bundle exists such
that the expected dimension of the moduli space is $d$, and $h$ is an element
of $H_2(X^4)$.

It is not clear that this series converges, but the convergence is not
crucial since the manipulations involved are formal.  Besides, in many
(perhaps all) cases the series does converge, giving rise to an honest
function $D:H_2(X)\to \re$.

Kronheimer and Mrowka say that a manifold $X^4$ satisfies the {\em
simple type} condition if $Q_{|z|+8}(x^2z)=4Q_{|z|}(z)$ for all $z\in
\re[x,y_1,\dots,y_{b_2}]$ and where $|z|$ means the degree of $z$.
This essentially says that the four-dimensional class $x$ is not
independent in the series $Q_d$, but satisfies $x^2=4$, so that these
invariants depend only on the $H_2(X^4)$ part.

When the simple type condition holds, then this series can be written as
\begin{equation}
D(h)=e^{I(h,h)/2}\cdot\left(r_1e^{K_1(h)}+\dots+r_me^{K_m(h)}\right)
\label{eqn:dseries1}
\end{equation}
where $I$ is the intersection form, viewed as bilinear function on
$H_2(X^4)$, $r_1, \dots, r_m$ are rational numbers, and $K_1, \dots,
K_m$ are elements in $H^2(X^4)$, thought of as linear functions on
$H_2(X^4)$.

They showed that for an impressive number of examples, the simple type
condition actually holds, and it has been conjectured that all
four-dimensional manifolds with $b_2^+>1$ are of simple
type.\footnote{There are examples, like $\cc P^2$, that are not of
simple type, but none of these have $b_2^+>1$.}  This appears likely,
especially in light of more recent developments.

In many cases, these series helped in the calculation of Donaldson
invariants, especially with examples from algebraic geometry.

\section{Algebraic and K\"ahler geometry}

When the manifold admits a K\"ahler metric, the anti-self-dual
equations to find instantons are related to the notion of stable
holomorphic bundles.\cite{AW,DK} Similarly, the self-dual equations
related to stable anti-holomorphic bundles.  Because mathematicians
prefer to deal with holomorphic objects rather than anti-holomorphic
objects, the orientation was chosen to deal with the anti-self-dual
equations.
\footnote{Essentially, if we view manifolds as coming with a
particular orientation (the fact that an orientation can be chosen at
all follows from the simply connected criterion), then we can view
solutions to the self-dual equations as solutions to the
anti-self-dual equations with the reverse orientation.  When the
orientation is reversed, $b_2^-$ becomes $b_2^+$, the sign convention
on $c_2(E)$ (as an integer) reverses, the intersection form becomes
its negative, and self-dual solutions become anti-self-dual solutions,
but nothing else changes.  So we can emphasize the theory of
anti-self-dual connections and discard the theory of self-dual
connections, without losing any real mathematics, as long as we
consider manifolds as coming with an orientation.}

But beyond this, this meant that for K\"ahler manifolds, there were a
number of ways of calculating the set of instantons without having to
actually solve a differential equation, and this led to important
advances in calculating Donaldson invariants, especially once
Kronheimer and Mrowka gathered them into the Donaldson
series.\cite{KM1,FS1}  For example, the K3 surface has Donaldson
series
\[\exp(I/2),\]
and more generally, an elliptic surface $E(\chi,m_1,\dots,m_r)$ (which has
elliptic curves as fibers and $S^2$ as base, where the holomorphic
Euler characteristic of this fibration is $\chi$ and $m_1, \dots, m_r$ are
the multiplicities of the singular fibers) has Donaldson series
\[\exp(I/2)\frac{(\sinh F)^{p_g-1+r}}{\prod_i \sinh(F/m_i)}\]
where $F$ is the class generated by the fiber and $p_g$ is the
geometric genus.\cite{FS1}  Similarly, many such formulas were computed
for complex algebraic surfaces of various kinds.

The main point here is that for complex algebraic surfaces, the
Donaldson invariants could be calculated systematically, once the
systematic framework of the Donaldson series of Kronheimer and Mrowka
was in place.  We will soon refer to the fact that the Donaldson
invariants of K3 are non-zero.

\section{Floer homology and topological quantum field theories}

One idea to calculate Donaldson invariants for non-K\"ahler manifolds is
to split the manifold into pieces, each of which might be easier to
analyze, perhaps because they can be viewed as pieces of K\"ahler manifolds.
This is easier said than done.  For an analogy, consider a large
molecule.  Strictly speaking, to predict the structure of the molecule,
you would need to solve a huge quantum $n$-body problem that is
beyond the capabilities of even the best computers.  But a more
enlightening approach might be to solve the Schrodinger equation for
individual atoms, then understand how orbitals combine to form
molecular orbitals in particular bonds, and so on.  Of course, you would
run into a few surprises that happen on a more global scale, but for the most
part, this idea works remarkably well.

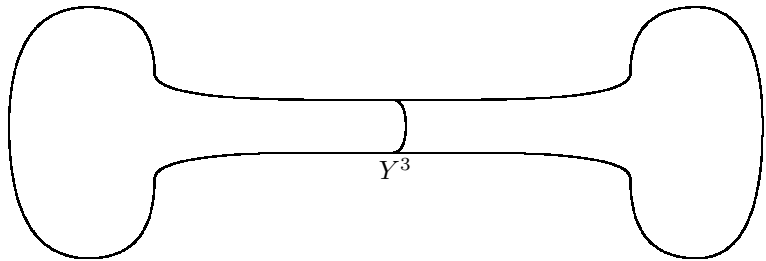
\begin{figure}
\begin{center}
\begin{picture}(300,95)(0,25)
\qbezier(5,75)(5,120)(35,120)
\qbezier(35,120)(60,120)(60,95)
\qbezier(60,95)(60,85)(120,85)
\qbezier(120,85)(150,85)(180,85)
\qbezier(180,85)(240,85)(240,95)
\qbezier(240,95)(240,120)(265,120)
\qbezier(265,120)(290,120)(290,75)
\qbezier(5,75)(5,25)(35,25)
\qbezier(35,25)(60,25)(60,55)
\qbezier(60,55)(60,65)(120,65)
\qbezier(120,65)(150,65)(180,65)
\qbezier(180,65)(240,65)(240,55)
\qbezier(240,55)(240,25)(265,25)
\qbezier(265,25)(290,25)(290,75)
\qbezier(150,85)(155,85)(155,75)
\qbezier(155,75)(155,65)(150,65)
\put(145,55){{$Y^3$}}
\end{picture}
\end{center}
\caption{Stretching the neck to calculate Donaldson invariants}
\label{fig:neck}
\end{figure}

For Donaldson theory, the idea would be to find a three-dimensional
submanifold that splits the four-dimensional manifold into two parts.
Since the Donaldson invariants are invariant under change of metric,
we can imagine altering the metric on the manifold by moving the two
parts further apart, which stretches a neighborhood of the three-dimensional
manifold into a neck (see Figure~\ref{fig:neck}).

Then instantons on the four-dimensional manifold can be viewed as
instantons on each piece which are glued together along instantons in
the cylindrical neck.  When we adjust the metric on the four-dimensional
manifold so that the neck becomes very long, the cylindrical
neck can be viewed as $Y^3\times \re$, where $Y^3$ is the
three-dimensional manifold.  The instanton solutions, from the
perspective of each unstretched side, become solutions that converge
to a constant solution on the $Y^3\times\re$ cylindrical end, while on
the neck itself, these solutions become solutions on the cylinder
$Y^3\times\re$.

A dimensional reduction can be done to turn the anti-self-dual equations
on the cylindrical neck $Y^3\times \re$
into a problem of connections $A$ on $Y^3$ (taking the temporal
gauge $A_0=0$).  Viewing $\re$ as time, the anti-self-dual equations become
the following flow:
\begin{equation}
\frac{d}{dt} A = -*F.
\label{eqn:floer}
\end{equation}

This equation is exactly the gradient flow equation in Morse theory.
Morse theory\cite{Ko} considers a smooth potential function $V:X^n\to \re$
and its critical points (points $p \in X^n$ so that $\nabla V(p)=0$).
The second-derivative test tells us whether the critical point is
a maximum, a minimum, or some kind of saddle in between.  More precisely,
the number of negative eigenvalues of the Hessian at $p$ is called the
index of the critical point $p$.  Minima have index $0$, and maxima
have index $n$.

Morse theory describes how the number of critical points of each index
relates to the homology of the manifold $X^n$.  Essentially, each
critical point of $V$ of index $k$ gives rise to a $k$-dimensional
cell in a decomposition of $X^n$ into cells, and these cells can be
used to compute the homology of $X^n$.  The trick is to look at
trajectories $x(t)$ in $X^n$ that satisfy
\[\frac{dx}{dt}=-\nabla V(x(t)).\]
If $p\in X^n$ is a critical point of $V$, then the set of points that
lie on trajectories that come from $p$ forms a cell whose dimension
is the index of $p$.  These cells, in turn, can be used to calculate
the homology of $X^n$, and in particular, give rise to formulas like
\[\sum_{i=0}^n (-1)^i b_i = \sum_{i=0}^n (-1)^i m_i\]
and
\[b_i\le m_i\]
where $m_i$ is the number of critical points of index $i$ and $b_i$
is the $i$th Betti number, that is, the rank of $H_i(X^n)$.

Given two critical points $p$ and $q$, there may be trajectories that
start from $p$ and end at $q$.  When the index of $p$ is one greater
than the index of $q$, there will be only finitely many such trajectories,
up to translation in $t$.  The number of these trajectories
can be used to reconstruct the homology of $X^n$, as long as the dimension
$n$ is finite.

Now replace $X^n$ with the set $\mathcal{A}$ of $SU(2)$ connections on
$Y^3$, up to gauge equivalence.  This is an infinite-dimensional
manifold.  For our potential function $V$ we take the Chern--Simons
functional
\[CS[A]=
\frac{1}{2}\int_{Y^3} \Tr\left( dA\wedge A + \frac{2}{3} A\wedge A\wedge
A\right)\]
Then
\[\frac{dA}{dt}=-\nabla CS[A(t)]\]
becomes
\[\frac{dA}{dt}=-*F\]
which is equation (\ref{eqn:floer}).  Analogously to the
finite-dimensional example, we can define a sort of ``homology'' by
counting flow lines between pairs of critical points of $CS$ whose
indices differ by one.  Because of the infinite dimensionality, the
result is not the homology of the configuration space, but it is
interesting nonetheless.  This is called the {\em Floer homology}
of $Y^3$.  There are a number of technical difficulties in carrying
this idea out, because the set of connections modulo gauge is
infinite-dimensional and because of the sort of non-compactness
associated with instantons, so that certain ideas of Morse theory are
no longer available and other techniques need to be invented.  But for
three-dimensional manifolds $Y^3$ with $b_1(Y^3)=0$, this idea was
carried out by A. Floer and refined by others (see, for instance,
Ref.~\cite{Fl1}).

The result is that the Donaldson invariants on a four-dimensional
manifold with $Y^3$ as boundary can be viewed as an element of the
Floer homology of $Y^3$; and when two four-dimensional manifolds have
the same boundary and are glued together along the boundary, an inner
product on the Floer homology pairs these invariants together to get
actual numbers (which, put together, form the Donaldson series).

This idea was explained by Sir Michael Atiyah\cite{At1}.  He views it
as a topological quantum field theory, where three-dimensional
manifolds are viewed as a possibility for a space-like slice, and
four-dimensional manifold that are cobordisms between these
three-dimensional manifolds are viewed as world-sheets.  The Hilbert
space of states associated to a space-like slice is the Floer homology
of the three-dimensional manifold, and the Donaldson invariant on the
four-dimensional manifold is the operator for time-translation.  This
is called a topological field theory since there is no local dynamics
(the Hamiltonian $H$ is zero).  The dynamics can only happen because
of topological changes in the space.  For a clear axiomatic treatment
of topological quantum field theories, see Ref.~\cite{At2} by Atiyah.

In the case where $X^4$ is a connected sum of two manifolds $X_+$ and
$X_-$, so that $X=X_+\#X_-$, we are attaching the two pieces along an
$S^3$.  Now $S^3\times\re$ is fairly easy to analyze, so in this case,
it is possible to show that if $b_2^+(X_+)$ and $b_2^+(X_-)$ are greater
than zero, then the Donaldson invariants for $X=X_+\#X_-$ vanish.

Therefore, since the Donaldson invariants were known not to vanish
for the K3 surface, we know that the K3 surface cannot be a connected
sum unless one of the pieces has $b_2^+=0$.  By Donaldson's Theorem A,
if this other piece has $E_8$'s, it must have at least one $H$, which
would make $b_2^+>0$; so if the K3 surface is a connected sum, it must
be with a piece that has the homology of a sphere.

Though there is a great deal of physics that underlies Donaldson theory, it
was possible for a mathematician to work in the field with little or
no knowledge of physics.  The reason is that although the physics
motivated the notion of gauge theory, it is possible to treat the
Yang--Mills theory as simply a minimization problem, and the self-dual
and anti-self-dual equations as partial differential equations.  This
is like learning, in an ordinary differential equations course, how to
solve second-order linear differential equations with constant
coefficients, without ever knowing about Hooke's law about springs.
In the case of finding instantons, there was a great deal of
interesting analysis, geometry, and topology that can be done without
knowing the physics, and once it was appreciated that on K\"ahler
manifolds these instantons related to stable holomorphic bundles, it
was possible to use algebraic geometry and complex geometry, too.
Atiyah's notion of a topological quantum field theory, also, could be
appreciated without even knowing what an ordinary quantum field theory
was, simply by thinking about what happened to instantons when you
stretch the neck.

Thus, many mathematicians went into the subject with little or no background
in quantum field theory, and were able to make important contributions.
The physics had its impact in posing a mathematical problem (the problem
of finding instantons) and from then on, mathematicians could play with
the problem without knowing the physics origins.  But there was also a
sense that if physics made an impact on mathematics in this unexpected
way once, perhaps it might happen again, and so ignoring the physics
would be a bad idea.  Perhaps physics might have more to say about
instantons and Donaldson invariants.  It did, as we will see next.

\section{Witten's work on Donaldson invariants}
Edward Witten found another way to understand Morse theory, relating
the cohomology of $X^n$ to critical points of a potential function
$V:X^n\to \re$.\cite{W1} He used a Hamlitonian formulation of a
certain $N=2$ supersymmetric quantum mechanical system.  He defined,
for every $t\ge 0$,
\[d_t=e^{-t V}de^{t V},\quad d_t^*=e^{t V}d^*e^{-t V},\]
then defined
\[Q_{1t}=d_t+d_t^*,\quad Q_{2t}=i(d_t-d_t^*),\quad H_t=d_td_t^*+d_t^*d_t,\]
and showed that for all $t$,
\[Q_{1t}^2=Q_{2t}^2=H_t,\quad \{Q_{1t},Q_{2t}\}=0.\]
The number of zero-modes of $H_0$ is the space of harmonic $k$-forms,
which has dimension $H^k(X^n)$.  As $t\to \infty$, the zero-modes of $H_t$
concentrate at the critical points of $V$, and it is possible to derive
the same kinds of relations between critical points of $V$ and $H^k$ that
show up in Morse theory.

As we saw in the previous section, Donaldson invariants on a cylinder
$Y^3\times\re$ is formally Morse theory for the Chern--Simons
functional.  So applying Witten's idea to Donaldson invariants on
$Y^3\times \re$ gives rise to an $N=2$ supersymmetric quantum system,
but since this Morse theory is on a set of gauge fields, we get a
quantum field theory.

Witten generated this quantum field theory in 1988\cite{W2}.  Roughly,
it involves gauge fields $A^a_i$ and anti-commuting fields
$\psi^a_i$ and $\chi^a_i$ on $Y^3$.  In the following discussion,
we will use the term {\em boson} to refer to commuting fields and
{\em fermion} to refer to anti-commuting fields, regardless of their spin.
Thus, $A^a_i$ is a boson, and $\psi^a_i$ and $\chi^a_i$ are fermions, even
though all three fields have spin 1.  This particular theory is not
relativistic, so the spin-statistics theorem does not apply.

The gauge group $SU(2)$ acts as usual on $A^a_i$ and on fermions
$\psi^a_i$ and $\chi^a_i$ it acts via the adjoint action.  The
Hamiltonian of this system is
\[H=\int\,d^3x\left[\frac{1}{2}\sum_{i,a}\left(-i\frac{\delta}{\delta A^a_i(x)}
\right)^2+\frac{t^2}{2}\Tr \tilde{F}_i \tilde{F}^i + t\epsilon_{ijk}
\Tr\psi^i D^j \chi^k\right].\]
The Latin index $i$ goes between 1 and 3, and represents space in
$Y^3$.

Roughly speaking, for each $i, a,$ and $x$, the $\psi_i^a(x)$ play the
role of a basis of one-forms on the set of connections $\mathcal{A}$,
and the $\chi_i^a(x)$ are a dual basis of vector fields on
$\mathcal{A}$.

Zero modes of the Hamiltonian will then calculate instantons on the
cylindrical manifold $Y^3\times \re$.  Since this theory was designed
for $Y^3$ and describes dynamics on $Y^3\times \re$, this cannot work
for a general four-dimensional manifold.  What was needed was an
extension of this theory into a four-dimensional covariant theory.

Witten then (in the same paper) came up with such a relativistically
covariant quantum field theory which was still supersymmetric, by
generalizing the existing fields and adding new fields.  In this
setting, $A^a_i$ became the gauge field $A^a_\mu$, where $\mu = 1, 2,
3, 4$ is a spatial 4-index; $\psi^a_i$ became $\psi^a_\mu$, and
$\chi^a_i$ became a self-dual two-form $\chi^a_{\mu\nu}$ (so that
$\tilde{\chi}_{\mu\nu}=\chi_{\mu\nu} =-\chi_{\nu\mu}$).  As before,
$A^a_\mu$ is bosonic and transforms as a connection under gauge
transformations, and $\phi^a_\mu$ and $\chi^a_{\mu\nu}$ are fermionic
and transform under the gauge group by the adjoint action.  There is
also a new fermionic scalar $\eta^a$.  Roughly speaking, the
$\psi_\mu$, $\chi_{\mu\nu}$ and $\eta$ fields occur because the
anti-self-dual equations, together with the gauge condition, to first
order are
\begin{eqnarray*}
(d+*d)A=0\\
*d(*A)=0
\end{eqnarray*}
and we can think of this as finding the zeros of an operator
$(d+*d,*d*)$ that sends one-forms to a pair of a self-dual two-form and a
scalar.  The $\psi^a_\mu(x)$, $\chi^a_{\mu\nu}(x)$, and $\eta^a(x)$ form
a basis for these spaces of forms.

Witten then added two bosonic scalar fields $\phi^a$ and $\rho^a$ to
balance the fermionic and bosonic degrees of freedom.  These also
transform via the adjoint action of the gauge group.

The commuting fields (bosons) are $(A_\mu, \phi, \rho)$ and the
anti-commuting fields (fermions) are $(\psi_\mu, \chi_{\mu\nu},
\eta)$.  Note that there is no relation between the spin and the
statistics (all the fields have integer spin), but this is explained
by the fact that some of these fields are ghosts.

The Lagrangian Witten discovered is
\begin{eqnarray}
\lagr&=&\int_{X^4} d^4x\sqrt{g}\left[
\frac{1}{4}F_{\mu\nu}F^{\mu\nu}
+\frac{1}{2}\phi D_\mu D^\mu\rho
-i\eta D_\mu\psi^\mu
+ i D_\mu\psi_\nu \chi^{\mu\nu}\right.\\
&&\left.-\frac{i}{8}\phi[\chi_{\mu\nu},\chi^{\mu\nu}]
-\frac{i}{2}\rho[\psi_\mu,\psi^\mu]
-\frac{i}{2}\phi[\eta,\eta]
-\frac{1}{8}[\phi,\rho]^2
+\frac{1}{4}F_{\mu\nu} \tilde{F}^{\mu\nu}\right]
\end{eqnarray}

The last term is a multiple of the second Chern class $c_2(E)$ and for
infinitessimal variations is irrelevant; but for later calculations it
is convenient.  Note that in Ref.~\cite{W2}, Witten calls $\lagr$ the
Lagrangian without this last term, and $\lagr'$ the Lagrangian
with it included.

If $(A,\phi,\rho,\eta,\psi,\chi)$ have scaling dimensions
$(1,0,2,2,1,2)$, this Lagrangian is scale invariant, and there is
another additive quantum number $U$ for which the fields have values
$(0,2,-2,-2,-1,1,-1)$, which is preserved by the Lagrangian.  This $U$
was a holdover from the Floer theory, where quantum-mechanical
violations of $U$ measured the degree in Floer homology.  In this new
setting, quantum-mechanical $U$ violation will turn out to measure the
dimension of the moduli space.  Furthermore there is a constant
spinless supersymmetry $Q$:
\begin{eqnarray}
&\delta A_\mu=i\epsilon \psi_\mu,\quad
\delta\phi=0,\quad \delta\rho=2i\epsilon\eta,&\\
&\delta \eta=\frac{1}{2}\epsilon[\phi,\rho],\quad
\delta\psi_\mu=-\epsilon D_\mu\phi,\quad
\delta\chi_{\mu\nu}=\epsilon(F_{\mu\nu}+\tilde{F}_{\mu\nu})&
\label{eqn:susy}
\end{eqnarray}
where $\epsilon$ is an antisymmetric scalar constant.  We write
$-i\epsilon\{Q,\obsv\}$ for $\delta \obsv$.  Then
$Q^2=0$ (though this fact uses the equations of motion in the case of $\chi$).

Again, the supersymmetry being a scalar is unexpected, but remember that
the spin-statistics theorem need not apply here.

This supersymmetric field theory may have been motivated by the Morse
theoretic view of instantons on $Y^3\times \re$, but Witten points out
that on flat $\re^4$, it is a ``twisted'' version of the usual $N=2$
supersymmetric gauge theory.

The usual $N=2$ SUSY theory with gauge fields and no matter fields has
the following multiplet structure (the ``$N=2$ gauge multiplet''):\cite{AH}

\begin{tabular}{ccc}
&$A_\mu^a$&\\
$(\lambda_{\alpha}^a,\bar{\lambda}_{\dot{\alpha}}^a)$&&$(\psi_{\alpha}^a,
\bar{\psi}_{\dot{\alpha}}^a)$\\
&$\varphi^a$&\\
\end{tabular}

The first row is the vector gauge field, the second row has two
fermionic spinors, and the bottom row is a complex bosonic scalar field.
There is a global internal symmetry $SU(2)_R\times U(1)_U$.  The
top and bottom rows are singlets for $SU(2)_R$, and the middle row
is an $SU(2)_R$ doublet.  $U(1)_U$ does not act on the gauge field,
it has charge 1 on $\lambda$ and $\psi$, and charge 2 on $\varphi$.

In Euclidean $(++++)$ space, $\lambda_{\alpha}$ and
$\bar{\lambda}_{\dot{\alpha}}$ are not complex conjugates and
are therefore separate fields.  So we really have four independent
fermionic spinors.  Also, $\varphi^a$ and $\varphi^{\dag a}$ are
two independent real fields that make up one complex bosonic scalar
field.

We will write our particle content, then, using the notation
$(n_-,n_+,n_R)^c$ where the numbers in the parentheses reflect the
representation of $SU(2)_-\times SU(2)_+ \times SU(2)_R$, and the
superscript is the representation of $U(1)_U$.

\begin{tabular}{llll}
field		&spin&statistics&
$(SU(2)_-\times SU(2)_+\times SU(2)_R)^{U(1)_U}$\\\hline
$A_\mu$		&$1$	&boson	&$(1/2,1/2,0)^0$\\
$(\lambda_\alpha,\psi_\alpha)$	
		&$1/2$	&fermion&$(1/2,0,1/2)^1$\\
$(\bar{\lambda}_{\dot{\alpha}},\bar{\psi}_{\dot{\alpha}})$
		&$1/2$	&fermion&$(0,1/2,1/2)^{-1}$\\
$\varphi$	&$0$	&boson	&$(0,0,0)^2$\\
$\varphi^\dag$	&$0$	&boson	&$(0,0,0)^{-2}$\\
\end{tabular}

Let $SU(2)'_+$ be the diagonal in $SU(2)_+\times SU(2)_R$.
Then $SU(2)_-\times SU(2)'_+\times U(1)_U$ is a symmetry.  We now
view $SU(2)_-\times SU(2)'_+$ as the spatial symmetry.  This is the ``twist''
we referred to.

Under $SU(2)_-\times SU(2)'_+\times U(1)_U$, the gauge fields $A_\mu$ will
not be affected, since there was no $SU(2)_R$ symmetry to begin with.  Its
representation is
\[(1/2,1/2)^0.\]
The spinor $SU(2)_R$ doublet $(\lambda_\alpha,\psi_\alpha)$ now acquires some
$SU(2)'_+$, and becomes a vector $\psi_\mu$:
\[(1/2,1/2)^1.\]
The other spinor $SU(2)_R$ doublet $(\bar{\lambda}_{\dot{\alpha}},
\bar{\psi}_{\dot{\alpha}})$ splits into two representations:
$\chi_{\mu\nu}$ and a boson $\eta$.
\[(0,1)^{-1}\oplus (0,0)^{-1}\]
The scalar bosons are unchanged, though we split $(\varphi,\varphi^\dag)$
into its real and imaginary parts, and call these $\phi$ and $\rho$.
\[(0,0)^2\oplus (0,0)^{-2}.\]
This is summarized in the following table:

\begin{tabular}{lllll}
field&spin&statistics&scale dim.&
$(SU(2)_-\times SU(2)'_+)^{U(1)_U}$\\\hline
$A_\mu$		&1&boson	&1&$(1/2,1/2)^0$\\
$\psi_\mu$	&1&fermion	&1&$(1/2,1/2)^1$\\
$\chi_{\mu\nu}$	&1&fermion	&2&$(0,1)^{-1}$\\
$\eta$		&0&fermion	&2&$(0,0)^{-1}$\\
$\phi+i\rho$	&0&boson	&0&$(0,0)^2$\\
$\phi-i\rho$	&0&boson	&0&$(0,0)^{-2}$\\
\end{tabular}

We note that these are precisely the particle fields in Witten's Lagrangian
above that is supposed to mimic Donaldson theory.  This is what is meant
when we say that Donaldson theory is a twisted $N=2$ SUSY theory.

There are two supersymmetries in the standard $N=2$ theory.  Under the
group $SU(2)_-\times SU(2)_+\times SU(2)_R \times U(1)_U$ the two
supersymmetries transform in the representation $(1/2,0,1/2)^{-1}$ and
$(0,1/2,1/2)^1$.  Under our twisted action of $SU(2)_-\times SU(2)'_+
\times U(1)_U$, the first one becomes $(1/2,1/2)^{-1}$ and the second
splits into $(0,1)^1\oplus (0,0)^1$.  The supersymmetry $Q$ we had
above was the $(0,0)^1$ component.

Since the usual $N=2$ theory is supersymmetric, so is Witten's twisted
theory, at least in $\re^4$ with the flat Euclidean metric.  What is
not clear is what happens when $\re^4$ is replaced by a more
general manifold, and Witten checks this and notes that Riemann
curvature considerations, which might normally appear, turn out not
to appear at all in this case.

Witten then describes, through a formal, non-rigorous argument, how
to compute the Donaldson invariants as expectation values (that is,
correlation functions) of the form
\[\langle W\rangle = Z(W) = \int (\DX) \exp(-\lagr/e^2) W\]
where $\DX$ indicates the integration over all the fields
$(A,\phi,\rho,\eta,\psi,\chi)$, $e$ is a constant viewed as a
``gauge coupling constant'', and $W$ is a polynomial in the fields
$A,\phi,\rho,\eta,\psi,\chi$.

We will not take an arbitrary $W$, but only those for which $\langle
W\rangle$ is invariant under changes in the metric, because otherwise
we won't have a topological invariant.

Before we proceed, we need a few basic facts.  The supersymmetry
of $\lagr$ and $\DX$ give rise to the equation $\langle
\{Q,W\}\rangle=0$.  Infinitessimal changes in the metric produce a
change in the Lagrangian by $\frac{1}{2}\int_{X^4} \sqrt{g} \delta
g^{\mu\nu}T_{\mu\nu}$, where $T_{\mu\nu}$ is the
stress-energy tensor.  Witten calculates this and shows that it is of
the form $T_{\mu\nu}=\{Q,\lambda_{\mu\nu}\}$ where
$\lambda_{\mu\nu}$ is an expression involving the fields that Witten
writes down explcitly.  Furthermore $\{Q,V\}=\lagr$ for some
expression $V$ in terms of the fields.

We will use these facts to find topological invariants.
For example, we will now show that the partition function $Z=Z(1)=\langle
1\rangle$ is invariant under changes in the metric.  To do this we
perturb the metric $g^{\mu\nu}$, and get:
\begin{eqnarray*}
\delta Z&=&\int(\DX)[\exp(-\lagr/e^2)]\cdot -\frac{1}{e^2}
\delta\lagr\\
&=&\int(\DX)[\exp(-\lagr/e^2)]\cdot -\frac{1}{e^2}
\frac{1}{2}\int_{X^4} \sqrt{g}\delta g_{\mu\nu}T_{\mu\nu}\\
&=&\int(\DX)[\exp(-\lagr/e^2)]\cdot -\frac{1}{e^2}
\frac{1}{2}\int_{X^4} \sqrt{g}\delta g_{\mu\nu}\{Q,\lambda_{\mu\nu}\}\\
&=&-\frac{1}{2e^2}\left\langle \left\{Q,\int_{X^4}\sqrt{g}\delta g^{\mu\nu}
\lambda_{\mu\nu}\right\}\right\rangle\\
&=&0
\end{eqnarray*}
so that $Z$ is a topological invariant.

The partition function $Z$ is also invariant under changes in $e$, as follows:
\begin{eqnarray*}
\delta Z&=&\int(\DX) \exp(-\lagr/e^2)\delta(-1/e^2)\lagr\\
&=&\delta(-1/e^2)\int(\DX) \exp(-\lagr/e^2)\{Q,V\}\\
&=&\delta(-1/e^2)\left\langle\{Q,V\}\right\rangle=0
\end{eqnarray*}

Therefore to calculate $Z$ we can take the limit of very small $e$, so
that the path integral is dominated by the classical minima.
The term depending on the gauge field $A$ is at a minimum whenever
$A$ is an instanton.  We can write the linearization of the anti-self-dual
equations, and the linearization of the gauge fixing condition, and
compute the number of degrees of freedom.  This is, of course, exactly the
index calculation earlier that gave the expected dimension of the
moduli space in (\ref{eqn:dim2}):
\[d=8c_2(E)-3(1-b_1(X^4)+b_2^+(X^4)).\]
The fermion zero modes turn out to give linearized equations that are
identical in form, but on $\psi_\mu$ instead of $A_\mu$.  Hence
the number of fermion zero modes equal the number of gauge zero modes.
It turns out there are generically no zero modes for $\eta$ and $\chi$.

Because of this, $Z$ vanishes, unless $d=0$.  This is precisely the
case where in classical Donaldson theory, we would want to count
points in the moduli space.  But this is exactly what this path
integral is doing.  The only thing to check is that the use of signs
agrees, which is delicate, but works.

Can we derive the other Donaldson invariants for $d>0$?  Yes.  What is
needed is the other expectation values $\langle W\rangle$.

We originally showed $Z$ is a topological invariant by varying the
metric.  The more general $Z(W)$ is a topological invariant when $W$
does not depend explicitly on the metric $g$ and when $\{Q,W\}=0$.
\footnote{The range of possible $W$ is slightly more general\cite{W2} but this
is not important for our purposes.}  So we need to find such
expressions.

If $W=\{Q,\obsv\}$, then although it is true that
$\{Q,W\}=0$, it is also true that
$Z(W)=\langle\{Q,\obsv\}\rangle=0$, so this does not help us.
Therefore, the set of expressions that we might use would be those $W$
(independent of $g$) for which $\{Q,W\}=0$, modulo those $W$ which are
$W=\{Q,\obsv\}$.  Thus, we are essentially looking for BRST singlet operators.

Upon examining the supersymmetry on the fields in equation (\ref{eqn:susy}),
we see that $\phi^a$ is invariant under $Q$, so that $\{Q,\phi^a\}=0$, and
$\phi^a$ is also not in the image of $Q$.  This means $\phi^a$ might
be a good candidate for $W$, except that $\phi^a$ is not
gauge invariant.

But $\Tr\phi^2$ is gauge invariant, and of course is still BRST singlet.  So
$W_0(P)=\Tr\phi^2(P)$ (where $P$ is a given point on $X^4$) is the
kind of functional we need, and $\langle W_0(P)\rangle$ is the
corresponding topological invariant.  It turns out not to depend on
$P$, as can be verified by differentiating with respect to $P$:
\[
\frac{\partial}{\partial x^\mu}W_0=
\frac{\partial}{\partial x^\mu}\left(\frac{1}{2}\Tr\phi^2(P)\right)
=\Tr\phi D_\mu\phi = i\{Q,\Tr \phi\psi_\mu\}\]
which has expectation value zero since it is of the form $\{Q,\cdot\}$.

Motivated by this, we define $W_1=\Tr(\phi\psi_\mu)dx^\mu$ as an
operator valued 1-form on $X^4$.  Thus,
\[0=i\{Q,W_0\}, \quad dW_0=i\{Q,W_1\}\]
Similarly, we can define $W_2, W_3, \dots$ like this:
\begin{equation}
dW_1=i\{Q,W_2\},\quad dW_2=i\{Q,W_3\},\quad
dW_3=i\{Q,W_4\},\quad dW_4=0
\end{equation}
and obtain the formulas
\begin{equation}
W_2=\Tr(\frac{1}{2}\psi\wedge\psi + i\phi\wedge F),\\
W_3=i\Tr(\psi\wedge F), \quad W_4=-\frac{1}{2}\Tr(F\wedge F).
\end{equation}
Then $W_k$ is an operator-valued $k$-form on $X^4$.

If $\gamma$ is a $k$-dimensional homology cycle on $X^4$, consider
$I=\int_\gamma W_k$.  We note that
\[\{Q,I\}=\int_\gamma \{Q,W_k\} = -i\int_\gamma dW_{k-1}=0,\]
so that $\langle I \rangle$ is a topological
invariant.  If $\gamma=\partial\beta$, then
\[I=\int_\gamma W_k=\int_\beta dW_k=i\int_\beta\{Q,W_{k+1}\}
=i\left\{Q,\int_\beta W_{k+1}\right\}\]
so that in that case, $\langle I\rangle=0$.  So $I$ only depends on
the homology class of $\gamma$.

So if $\gamma_1, \dots, \gamma_r$ are homology classes in degree
$k_1, \dots, k_r$, then
\[\left\langle\int_{\gamma_1} W_{k_1}\dots \int_{\gamma_r} W_{k_r}\right\rangle\]
is a topological invariant.  This is zero unless
\[\sum_{i=1}^r (4-k_r)=d\]
where $d$ is the expected dimension of the moduli space.  Witten then
adjusts $e$ to be small, as before, and relates it to the notion of
integrating over the moduli space $\moduli$ differential forms that
are canonically defined in $\bsp$.  This leads differential forms are,
essentially, $\mu(\gamma_i)$, and so these correlation functions turn
out to be, indeed, the Donaldson invariants $Q(\gamma_1\dots\gamma_r)$.

There is much that is unusual about this theory.  But one that is
striking is that what appeared to be a classical problem of
counting classical field theory solutions to the anti-self-dual
equation (finding instantons) turns out the be expressible as
correlation functions in a supersymmetric quantum field theory.

Although this formulation caught the attention of many, this approach
did not lead to many mathematicians using this approach to prove
theorems about Donaldson invariants and about four-dimensional
manifolds, perhaps for three reasons: first, it takes time and effort
for mathematicians trained in analysis and topology to learn the
relevant physics; second, the mathematics needed to make the physics
rigorous was not (and still is not) available, and the problem of
making all the arguments rigorous seems daunting; and third, it was
not clear whether or not these physical insights could lead to new
theorems, or even lead to new explicit calculations.  In 1994, Witten
showed how to calculate Donaldson invariants for K\"ahler
manifolds\cite{W3}, but this was just already discovered by the work
of Kronheimer and Mrowka\cite{KM1} using other techniques (Witten
refers to these developments in his paper), and it was not clear
whether or not the non-physics-related methods could do everything
these physics-related methods could do.

But also in 1994, Witten's supersymmetric theory proved itself useful
again, this time in a far more dramatic way.

\section{Seiberg--Witten theory and $S$ duality}
In 1988, Nathan Seiberg had discovered new techniques to show that
certain supersymmetric theories had very explicit formulas for quantum
corrections, obtained by considerations of holomorphicity and
symmetry.\cite{S0} The idea is that if the quantum theory has
supersymmetry, this constrains the form of the effective Lagrangian so
much that it is possible to describe to write down explicit formulas,
at which point it so happens that the quantum corrections vanish after
the one-loop stage to all orders in perturbation theory.  Instantons
give rise to non-perturbative effects, but the form of this is highly
constrained, too, so that it is possible to derive explicit formulas
for the effective Lagrangian.

In 1994, Nathan Seiberg and Edward Witten\cite{S1,S2} used these
techniques to discover new dualities in supersymmetric theories, and
was able, in a short amount of time, to illustrate many features of
supersymmetric gauge field theories that had eluded physicists for
decades for more general theories (like quark confinement)\cite{SW1}.
These dualities exchange electric and magnetic charges, while turning
a weak coupling constant into a strong one by $g\leftrightarrow 1/g$.

Dualities of this sort were first conjectured for gauge theories by
Olive and Montonen\cite{MO}, and were verified for $N=4$ by Olive and
Witten.\cite{WO} Seiberg and Witten were able to put this result in a
broader framework of dualities for more general supersymmetric
theories.  In other supersymmetric theories, the duality related two
different descriptions of the same theory.  These ideas led to the
current program of the unification of string theories into $M$-theory,
and many other exciting developments in supersymmetric theories in
various dimensions.  A brief description of this idea was reviewed by
Seiberg recently in this journal\cite{S2}, and an introduction for
beginners was written by Luiz Alvarez-Gaum\'e and
S. F. Hassan.\cite{AH}

The duality for pure $N=2$ supersymmetric gauge theory was worked out
in detail by Seiberg and Witten\cite{SW1,SW2}.  We have just seen that
the $SU(2)$ supersymmetric gauge theory can be twisted into a
topological field theory, for which the Donaldson invariants are the
correlation functions of BRST singlet operators.  The dual theory also
has $N=2$ supersymmetry, and so the same twist can be applied and we
will get a new, dual, topological field theory.  We would then expect
that the correlation functions of the BRST singlet operators in this
dual field theory should also be the Donaldson invariants.  This idea
was first described in a 1994 talk by Witten\cite{W4}.  The ideas were
made more precise and explicit by Moore and Witten.\cite{MW}

In the remainder of this section, I will derive the dual topological
field theory following the arguments of Seiberg and Witten.

First, let us review a few facts about $N=2$ SUSY Yang--Mills theory.
The gauge multiplet

\begin{tabular}{ccc}
&$A_\mu^a$&\\
$(\lambda_{\alpha}^a,\bar{\lambda}_{\dot{\alpha}}^a)$&&$(\psi_{\alpha}^a,
\bar{\psi}_{\dot{\alpha}}^a)$\\
&$\varphi^a$&\\
\end{tabular}

\noindent
can be viewed in $N=1$ superspace terms as a chiral superfield
$\Phi=(\varphi,\psi_\alpha)$ and a vector superfield $V=(A_\mu,\lambda_\alpha)$;
or in $N=2$ superspace terms as a gauge superfield $\Psi=(\varphi,\psi_\alpha,
A_\mu,\lambda_\alpha)$.  The renormalizable Lagrangian can be written as
\begin{equation}
\lagr=\frac{1}{4\pi} \im\Tr\int d^2\theta d^2\tilde{\theta}
\frac{1}{2}\tau \Psi^2
\end{equation}
where $\tau=\theta/2\pi+4\pi i/e^2$.  When we wish to find the
effective Lagrangian, if it still has $N=2$ supersymmetry then we
should expect the effective Lagrangian to have the same form, except
that since we do not need this to be renormalizable we can replace
$\Psi^2$ by any holomorphic function $\mathcal{F}(\Psi)$.  This
function $\mathcal{F}$ is called the prepotential.
\begin{equation}
\lagr=\frac{1}{4\pi} \im\Tr\int d^2\theta d^2\tilde{\theta}
\mathcal{F}(\Psi)
\label{eqn:n2lagr}
\end{equation}

Now Seiberg\cite{S0} calculates the prepotential $\mathcal{F}$ using
basic facts about the symmetries, and obtains the form
\begin{equation}
\mathcal{F}(\Psi)=\frac{i}{2\pi}\Psi^2\ln\frac{\Psi^2}{\Lambda^2}
+\sum_{k=1}^\infty \mathcal{F}_k\Lambda^{4k}\Psi^{2-4k}
\label{eqn:exact}
\end{equation}
where $\Lambda$ is a fixed dynamically generated scale.  The first
term is the one-loop correction, and there are no other perturbative
corrections.  The sum is due to non-perturbative ``instanton''
corrections, and the coefficients $\mathcal{F}_k$ are only known for small
values of $k$ on $\re^4$.

In terms of the $N=1$ fields $\Phi$ and $V$, we can write the Lagrangian as
\begin{equation}
\frac{1}{4\pi}\im\left[\int d^4\theta\,
\frac{\partial \mathcal{F}}{\partial \Phi}\overline{\Phi}
+\int d^2\theta\,
\frac{1}{2}\frac{\partial^2\mathcal{F}}{\partial \Phi^2}W_\alpha W^\alpha
\right]
\label{eqn:n1lagr}
\end{equation}
where $W$ is the field strength of the vector superfield $V$.
Furthermore the metric can be written as $\im \frac{\partial^2
\mathcal{F}}{\partial \varphi^2}$, and the coupling constant is given by
\[\tau=\frac{\partial^2\mathcal{F}}{\partial \varphi^2}.\]

One feature of the $N=2$ theory is a Higgs-like classical vacuum.  The
bosonic terms in the Lagrangian, after eliminating auxiliary fields,
are
\begin{equation}
\lagr_{boson}=
\frac{1}{e^2}\Tr\left(-\frac{1}{4} F_{\mu\nu}F^{\mu\nu}
+e^2\frac{\theta}{32\pi^2}F_{\mu\nu}\tilde{F}^{\mu\nu}
+(D_\mu \varphi)^\dag D^\mu\varphi - \frac{1}{2}[\varphi^\dag,\varphi]^2\right).
\end{equation}
There is a Higgs potential here $[\varphi,\varphi^\dag]$.  We have a
classical vacuum when $\varphi$ is a covariantly constant scalar field,
such that $[\varphi^\dag,\varphi]=0$.  This happens when the real and
imaginary parts of $\varphi$ point in the same direction.
After a gauge transformation to point the real part of $\varphi$ in
the direction $\sigma_3$, we end up with $\varphi=\frac{a}{2}\sigma_3$ for
$a\in \cc$.  Classically $a$ can take any value.  This turns out to be
true also in the full quantum theory.

In order to describe the classical moduli space of vacua, we cannot
use $\varphi^a$ since that is not gauge-invariant, so we use
$W_0=\Tr(\varphi^2(P)) =\frac{1}{2}a^2\in \cc$ to parameterize the
classical moduli space.  The set of vacua here (the complex plane), as
with classical instantons, is a family that is not generated by
symmetries, and in fact the effective theory localized around each of
these vacua is in general different.

For $W_0\not=0$, the $SU(2)$ gauge group is broken to $U(1)$.
Localized around this point in the moduli space, we have classical
monopole and dyon solutions.  These satisfy the
Bogomol'nyi--Prasad--Sommerfield mass relation
\[M=a\sqrt{g^2a^2n_e^2+\frac{a_D^2}{g^2}n_m^2}\]
where $n_e$ and $n_m$ are integers.

Seiberg takes the exact prepotential $\mathcal{F}$, and shows that
the quantum moduli space of vacua is this same degenerate moduli
space, and that $u=\langle W_0(P)\rangle\in \cc$ parametrizes this
moduli space of quantum vacua.  The situation where $u$ is large is
where coupling becomes weak, and the classical approximation in valid.
So here, $u\cong \frac{1}{2}a^2$, and the theory is singular at
$u=\infty$.  Now $a$ cannot be a good parameter, or else the metric,
given by the harmonic function
\[\im\frac{\partial^2 \mathcal{F}}{\partial a^2}\,da\,d\bar{a}\]
would eventually be negative by the maximum principle of harmonic functions.

As we move $u$ in a large circle where the theory is classical, we
see that since $u=\frac{1}{2}a^2$, $a$ will go to $-a$.  If we let
\[a_D=\frac{\partial \mathcal{F}}{\partial a}\]
which by (\ref{eqn:exact}) is
\[a_D=\frac{2ia}{\pi}\ln(a/\Lambda)
+\frac{ia}{\pi}+\sum_{k=1}^\infty (2-4k) F_k \Lambda^{4k} a^{1-4k}\]
then we see that $a_D$ goes to $-a_D+2a$.  This can be viewed as a
monodromy
\[{a_D\choose a}\to \left(\begin{array}{cc}
-1&2\\
0&-1
\end{array}\right){a_D\choose a}.\]
If there is a monodromy like this at $u=\infty$, there must be other
singularities for finite $u$, where the classical description is no
longer valid.  Furthermore, some monodromies must fail to commute with
the monodromy above, or else $a$ would be a good parameter, and the
metric would fail to be positive.  Therefore, there must be at least
two singularities in the finite $u$ plane.

There is a symmetry $u\to -u$, so it makes sense to suppose there are
precisely two singularities, which after rescaling the $u$ plane, may
be at $\pm 1$.  The situation $u=0$ would classically give rise to a
singularity, but this singularity no longer occurs quantum
mechanically.

There are many strong indications that there are only two singular
points in the finite $u$ plane, ranging from checking particular cases
explicitly, to calculating $\mathcal{F}_1$ and comparing the results
with those already known by explicit calculation, but a rigorous proof
is still lacking.  Seiberg and Witten show that the monodromies around
$\pm 1$ multiply to the monodromy at $\infty$, giving further
credibility to the idea that there are only two singularities in the
finite $u$ plane.  The fact that this duality gives rise to a theory
mathematicians are interested in might be considered as further
evidence, as I will describe later.

The monodromies generate $SL_2(\zz)$, and in particular include
\[S=\left(\begin{array}{cc}
0&1\\ -1&0\end{array}\right)\] which sends $a$ to $a_D$ and $a_D$ to
$-a$.  It also sends the coupling constant to its reciprocal.  This is
the $S$ duality mentioned above.  Also, $a$ and $a_D$ have the
following interpretation: a dyon will have electric charge $an_e$ and
magnetic charge $a_D n_m$, where $n_e$ and $n_m$ are integers.  They
will satisfy the BPS bound, $M=\sqrt{g^2(an_e)^2+(a_D n_m)^2/g^2}$.

The only immediately recognizable source of the singularities in the
$u$ plane is the possibility that the dyons become massless.  This
occurs when $a_D=0$ and $n_e=0$, so that these dyons are magnetic
monopoles.

The theory around these points $u=\pm 1$, therefore, appears to involve
massless magnetic monopoles, which are described as an $N=2$ $U(1)$ gauge
multiplet

\begin{tabular}{ccc}
&$A_\mu$&\\
$(\lambda_{\alpha},\bar{\lambda}_{\dot{\alpha}})$&&$(\psi_{\alpha},
\bar{\psi}_{\dot{\alpha}})$\\
&$\varphi$&\\
\end{tabular}

\noindent
and an $N=2$ $U(1)$ hypermultiplet

\begin{tabular}{ccc}
&$\xi_\alpha$&\\
$\Phi$&&$\tilde{\Phi}$\\
&$\tilde{\xi}_{\dot{\alpha}}$&\\
\end{tabular}

\noindent
where $\xi$ and $\tilde{\xi}$ are spinors that are fixed by the $SU(2)_R$
symmetry, and $\Phi$ and $\tilde{\Phi}$ are scalars that fit into one $SU(2)_R$
doublet.  Near $u=\pm 1$, the low energy theory is dominated by this
behavior, with coupling constant $-1/\tau$ instead of $\tau$.  At the
actual singularity, $\tau$ goes to infinity, but in terms of the monopoles,
the coupling constant goes to zero, and we can use a semiclassical
approximation.

This is the dual theory to $N=2$ supersymmetric gauge theory.

\section{Seiberg--Witten duality on Donaldson theory}

As we saw before, Donaldson theory is a twisted version of $N=2$ SUSY
pure gauge theory, so we might apply the above to Donaldson theory.

Recall the relationship between Donaldson invariants and $N=2$ SUSY
pure gauge theory.  Starting from $N=2$ SUSY pure gauge theory,
and using the Euclidean $++++$ metric, we ``twist'' the
$\spin(4)=SU(2)_-\times SU(2)_+$ with the $SU(2)_R$ and obtain
a topological quantum field theory where certain correlation coefficients
do not depend on the metric or the coupling constant $e$.

Consider a one-parameter family of metrics $g_t=t^2g_1$ for some fixed
metric $g_1$.  If $t\to 0$, we get the weak coupling limit $e\to 0$,
and the theory is dominated by the $A_\mu$ fields and their minima,
which are the classical instantons.  In terms of the $u$ plane
description, as the coupling goes to zero, the region in the $u$ plane
where the theory is classical (near $u=\infty$) expands to include
more and more of the $u$ plane, until in the limit, the contribution
only comes from $u=0$, where classically the full $SU(2)$ gauge theory
is unbroken.

Since the correlation functions do not depend on the metric or the
gauge coupling, we can compute the same things as $t\to \infty$.  In
this direction, we no longer have a classical theory, and it appears
we should integrate over the $u$ plane.  But it turns out that for
most of the $u$ plane, there are no contributions, as long as
$b_2^+>1$.  The reason is roughly that there are too many fermionic zero-modes
that force the contribution to be zero away from $u=\pm 1$.

So just as in a contour integral in complex analysis, the
contributions only come from $u=\pm 1$.  At this point, we can use
the dual description of the theory in terms of massless monopoles, and
in this description, the coupling constant is small.

We apply the Witten twist again, and this time we get not only the
gauge multiplet (with $SU(2)$ broken to $U(1)$) but a hypermultiplet
describing the monopole.  The gauge multiplet twists in exactly the
same way as in the Donaldson case above.  The hypermultiplet goes from

\begin{tabular}{lllll}
field		&spin&statistics&$SU(2)_-\times SU(2)_+\times SU(2)_R$\\\hline
$\xi_\alpha$	&$1/2$&	fermion	&$(1/2,0,0)$\\
$\Phi$,$\tilde{\Phi}$	&$0$&	boson	&$(0,0,1/2)$\\
$\tilde{\xi}_{\dot{\alpha}}$
		&$1/2$&	fermion	&$(0,1/2,0)$
\end{tabular}

\noindent
to

\begin{tabular}{lllll}
field		&spin&statistics&$SU(2)_-\times SU(2)'_+$\\\hline
$\xi_\alpha$	&$1/2$&	fermion	&$(1/2,0)$\\
$\Phi_\alpha$	&$1/2$&	boson	&$(0,1/2)$\\
$\tilde{\xi}_{\dot{\alpha}}$
		&$1/2$&	fermion	&$(0,1/2)$
\end{tabular}

\noindent
and we see that we get only spinors, one of which is a boson.  We take
the limit as the ``new'' coupling constant goes to zero, and we get
a theory dominated by $A_\mu$ as before but also the spinor $\Phi_\alpha$.
The correlation functions are dominated by solutions to a certain
classical equation of motion, which we describe in the next section.

This new classical theory is what mathematicians have called {\em
Seiberg--Witten theory}.  The use of this term led to some confusion
when discussing these developments with physicists, who were used to
using this term for the general approach to supersymmetry that these
techniques inspired.  Thus, mathematicians were in the unusual
situation of using a term in a more restrictive sense than physicists.

\begin{figure}[t]
\begin{sideways}
\begin{picture}(480,350)
\put(0,175){{$t$}}
\put(10,300){\vector(0,-1){250}}
\put(10,303){\makebox[0in]{$\infty$}}
\put(10,40){\makebox[0in]{$0$}}
\put(20,175){{$e$}}
\put(30,300){\vector(0,-1){250}}
\put(30,303){\makebox[0in]{$\infty$}}
\put(30,40){\makebox[0in]{$0$}}
\put(40,160){\framebox[1in]{\parbox[b]{1in}{\begin{center}
Witten--Donaldson\\
$(A_\mu,\psi_\mu,\chi_{\mu,\nu},$\\
$\rho,\phi)$
\end{center}}}}
\put(40,0){\framebox[1in]{\parbox[b]{1in}{\begin{center}
Donaldson\\
$SU(2)$ $A_\mu$\\
$*F=-F$\end{center}}}}
\put(75,150){\vector(0,-1){86}}
\put(130,175){\vector(-1,0){20}}
\put(150,255){\vector(-1,0){60}}
\put(120,255){\makebox[0in]{\parbox[b]{2in}{\begin{center}
Witten twist
\end{center}}}}
\put(130,150){\framebox[1in]{\parbox[b]{1in}{\begin{center}$N=2$ SUSY\\
Pure $SU(2)$\\
$A_\mu$\\
$\lambda_\alpha$\quad $\psi_\alpha$\\
$\varphi$\end{center}}}}
\put(202,175){\vector(1,0){18}}
\put(220,175){\vector(-1,0){18}}
\put(172,130){\vector(1,0){78}}
\put(250,130){\vector(-1,0){78}}
\put(211,130){\makebox[0in]{\parbox[t]{2in}{\begin{center}
$S$ duality\\
$e\leftrightarrow \frac{1}{e}$
\end{center}}}}

\put(220,145){\framebox[1.3in]{\parbox[b]{1.3in}{\begin{center}$N=2$ SUSY\\
$U(1)$ gauge and hypermultiplet\\
\parbox{.5in}{\begin{center}
$A_\mu$\\
$\lambda_\alpha$\quad $\psi_\alpha$\\
$\varphi$\end{center}}
\parbox{.5in}{\begin{center}
$\xi_\alpha$\\
$\Phi$\quad $\tilde{\Phi}$\\
$\tilde{\xi}_{\dot{\alpha}}$\end{center}}
\end{center}}}}
\put(314,175){\vector(1,0){16}}
\put(292,255){\vector(1,0){60}}
\put(322,255){\makebox[0in]{\parbox[b]{2in}{\begin{center}
Witten twist
\end{center}}}}
\put(330,155){\framebox[1.6in]{\parbox[b]{1.6in}{\begin{center}
$U(1)$ monopoles\\
\parbox{.5in}{\begin{center}
$A_\mu$\\
$\lambda_\mu, \chi_{\mu\nu},\eta$\\
$\varphi$\end{center}}
\parbox{.8in}{\begin{center}
$\xi_\alpha$\\
$\Phi_\alpha$\\
$\tilde{\xi}_{\dot{\alpha}}$\end{center}}
\end{center}}}}
\put(388,235){\vector(0,1){60}}
\put(330,300){\framebox[1.6in]{\parbox[b]{1.6in}{\begin{center}
Seiberg--Witten equations\\
$U(1)$\\
$A_\mu$, $\Phi_\alpha$
\end{center}}}}
\put(470,50){\vector(0,1){247}}
\put(480,175){$e$}
\put(470,40){\makebox[0in]{$\infty$}}
\put(470,300){\makebox[0in]{$0$}}
\end{picture}
\end{sideways}
\caption{The relationship between Donaldson theory, counting
instantons, on the bottom left, and the dual theory, described in
detail in the next section, on the top right, counting
monopoles.}
\end{figure}
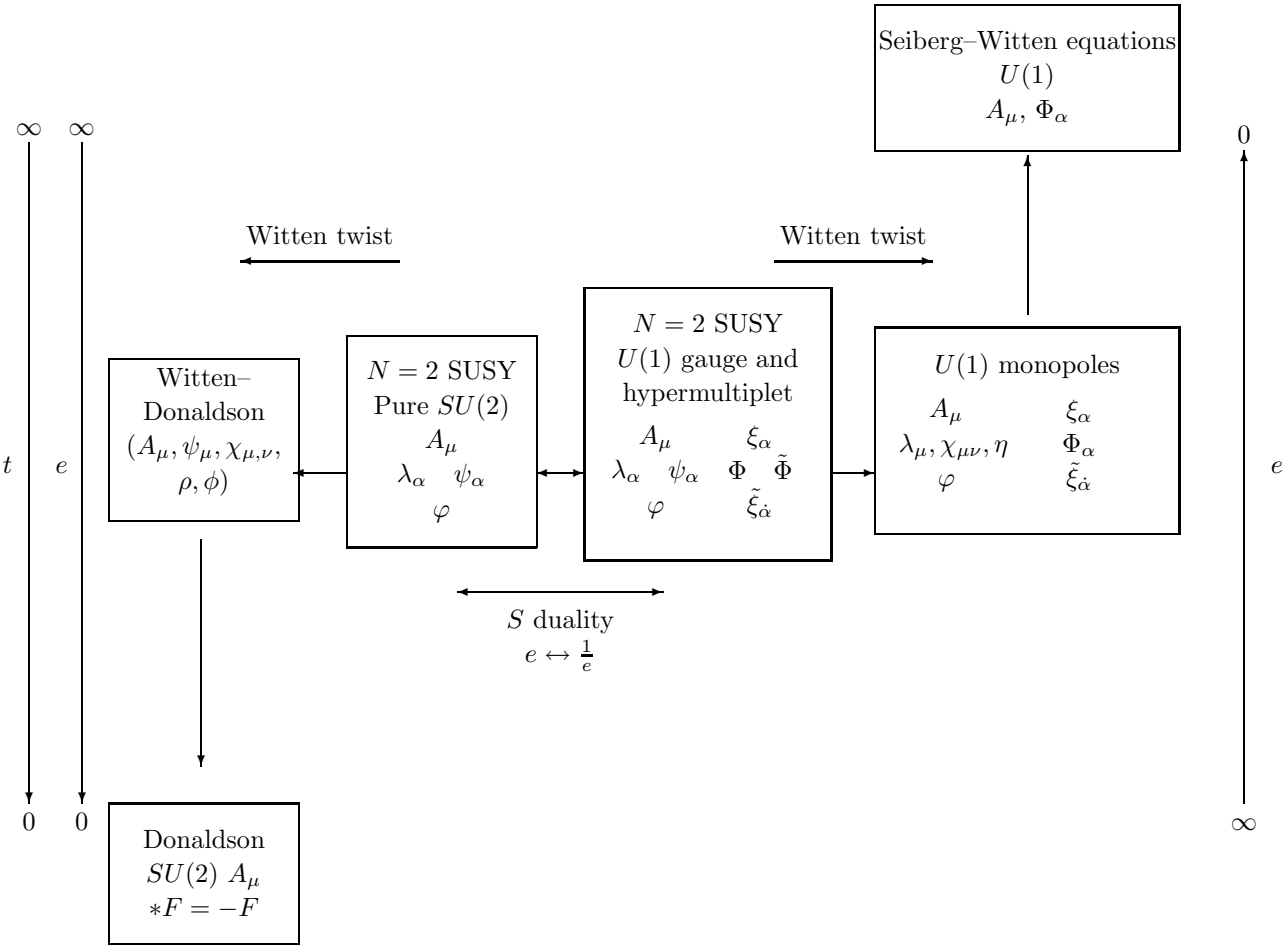

These results are explored more fully in Moore and Witten's
paper\cite{MW}, and the precise relationship between Donaldson theory
and this dual theory is also explained.

\section{The Seiberg--Witten equations}

Recall that for Donaldson invariants, we considered the moduli space
of instantons, that is, classical $SU(2)$ gauge-fields that satisfy
the self-dual or anti-self-dual equations, on a four-dimensional
manifold with positive definite metric.  The topology of the moduli
space inside the set of all connections modulo gauge gives rise to the
Donaldson invariants.

In this dual Seiberg--Witten theory, we are again considering the
moduli space of classical solutions to a differential equation coming
from a gauge theory, on a four-dimensional manifold with
positive-definite metric.  We again consider the topology of the moduli
space inside the set of all possible fields modulo gauge, and we hope
to define Seiberg--Witten invariants.

The gauge group $SU(2)$ is replaced by $U(1)$, though in addition to
the connection $A_\mu$ we also have the bosonic Weyl spinor $\Phi_\alpha$.
The vector bundle $E$ is now replaced by a complex line bundle $L$,
in accordance with $SU(2)$ breaking to $U(1)$.

Let us consider what sort of object $\Phi_\alpha$ is.  It arises from
the Witten twist on two scalars $\Phi$ and $\tilde{\Phi}$.  These
scalars lie in representations of $U(1)$.  Therefore, $\Phi_\alpha$
transforms under $\spin(4)\times U(1)$.  Actually, under this action
$(-1,-1)\in \spin(4)\times U(1)$ acts trivially, so the symmetry group
of $\Phi_\alpha$ is $\spin^c(4)=(\spin(4)\times U(1))/\pm 1$ where the
$\pm 1$ acts diagonally on both factors.  By projecting onto the first
factor, we have a homomorphism $\pi_1:\spin^c(4)\to \spin(4)/\pm
1=SO(4)$.  One way to describe the configuration $\Phi$ is to view it
as a section of a two-dimensional complex vector bundle $W$ over $X^4$
with a structure group $\spin^c(4)$ so that the projection to $SO(4)$
gives the same $SO(4)$ bundle as the tangent bundle.  There is
similarly a projection $\pi_2:\spin^c(4)\to U(1)/\pm 1\cong U(1)$
on the second factor.  It is useful to think of the $\spin^c(4)$ bundle
as having these two components: the first component is an $SO(4)$ bundle
that is the tangent bundle, and the second component is a $U(1)$ bundle
that corresponds to the gauge freedom in the theory.

The group $\spin^c(4)$ is the complex analogue of $\spin(4)$ in the
sense that it comes from the complexified Clifford algebra in the same
way as $\spin(4)$ comes from the Clifford algebra.  There are many
similarities between the two, including the existence of gamma
matrices and (given a $U(1)$ connection for the $U(1)$ piece) a Dirac
operator.

For $\spin(4)$, we know that the tangent bundle to a manifold $X^4$
has structure group $\spin(4)$ if and only if a certain
Stiefel-Whitney class $w_2$ is equal to zero.  For $\spin^c(4)$, it is
possible to work out the analogous conditions, and in four dimensions,
the tangent bundle always has a $\spin^c(4)$ bundle, without further
conditions.  There are typically many ways to make it a $\spin^c(4)$
bundle, parameterized by the $U(1)$ part.  One way to extract the
$U(1)$ information from $W$ is to construct a complex line bundle over
$X^4$ so that the gauge transformations on the line bundle are given
by the determinants of gauge transformations on $W$.  This complex
line bundle is called $\det(W)$.  Taking the first Chern class
$c_1(\det(W))$ gives a class in $H^2(X^4)$.  It turns out that this
class in $H^2(X^4)$ completely classifies the $\spin^c(4)$ bundle.

Then the the Seiberg--Witten equations are
\begin{eqnarray}
F_{\mu\nu}+*F_{\mu\nu}
&=&-\frac{i}{2}\bar{\Phi}^{\dot{\alpha}}
\Gamma_{\mu\nu\dot{\alpha}\beta}\Phi^\beta\\
\Gamma^\mu_{\dot{\alpha}\beta} (D+A)_\mu \Phi^\beta&=&0.
\label{eqn:sw}
\end{eqnarray}

Here $F_{\mu\nu}$ is the curvature of $A_\mu$, and
$\bar{\Phi}\Gamma_{\mu\nu}\Phi$ lies naturally in
$\Lambda^{2,+}T^*X^4$, which makes the first equation sensible.  The
second equation is, of course, the Dirac equation, which requires that
we use the $U(1)$ connection $A$ together with the Levi-Civita $\spin$
covariant derivative $D$ to get the $\spin^c(4)$ covariant derivative
$(D+A)_\mu$.

Also recall that $A_\mu$ and $\Phi_\alpha$ are both $c$-numbers, even
though $\Phi_\alpha$ is a spinor.

As in the instanton case, we need to find the set of classical
solutions to equations (\ref{eqn:sw}), up to gauge equivalence.  As
before, it is possible to prove that the set of solutions forms a
manifold of dimension
\begin{equation}
d=\frac{1}{4}\left(c_1(\det(W))^2-(2+2b_1+4b_2^+-2b_2^-)\right).
\label{eqn:swdim}
\end{equation}

So we have another theory that seems to be similar to Donaldson
theory, at least in the basic features.  Of course, this development
would not have been a revolution in our understanding of
four-dimensional manifolds unless it were a simplification of the
anti-self-dual equations of Donaldson theory, and so it is.  The fact
that the gauge group has gone from $SU(2)$ to what is essentially
$U(1)$ might be expected to make matters much simpler, but this is
somewhat of a red herring.  The biggest reason why abelian theories
are easier than non-abelian ones is because the curvature is a linear,
rather than non-linear, operator on the set of connections.  But the
fact that we have gone from a non-abelian gauge theory to an abelian
one is offset by the fact that we have two quadratic non-linearities that
appear: $\bar{\Phi}\Gamma \Phi$ in the first equation, and the
interaction between $A$ and $\Phi$ because of the second equation.

In other words, we may not have self-interactions in the gauge field, but
we have interactions between the gauge field and the matter field in two
different ways, and this amounts to the same kind of problems.

There is, however, an important and useful feature of these
Seiberg--Witten equations that is absent in Donaldson theory: the set
of solutions is bounded by the scalar curvature.  The reason for this
is we have a Weitzenb\"ock formula
\[
(\Gamma (D+A))^*(\Gamma (D+A))\Phi
=\nabla_A^*\nabla_A\Phi+\frac{R}{4}\Phi-\frac{1}{2}F^+ \Gamma \Phi
\]
where $R$ is scalar curvature, and $\Gamma$ is assumed to carry the
indices to contract completely with the curvature $F$ or the
covariant derivative $D$, as the context suggests.  Note that although
$R$ is not necessarily constant on $X^4$, we assume that $X^4$ is
compact, so that the values of $R$ are bounded on $X^4$.

Consider a configuration $(A_\mu,\Phi_\alpha)$ that satisfies the
Seiberg--Witten equations (\ref{eqn:sw}).  At a point $P\in X^4$ where
$\Phi$ is at a maximum, we note that $\Delta |\Phi|^2\le 0$.  We then use
the Weitzenb\"ock formula to rewrite the Laplacian in terms of the
Dirac operators.  Using the equations of motion we can derive that at
such a maximum point $P\in X^4$,
\[-\frac{R(P)}{2}|\Phi(P)|^2-\frac{1}{2}|\Phi(P)|^4\ge 0\]
which shows that either $|\Phi(P)|^2=0$ or $|\Phi(P)|^2=-R(P)$.
Of course, at the point $P$, $|\Phi(P)|$ was assumed to be a maximum,
so if $|\Phi(P)|^2=0$, then $\Phi$ is zero everywhere on $X^4$.  Otherwise
if the scalar curvature is bounded from below by some number $k$ (which it
must be if $X^4$ is compact), then $R\ge k$, and $|\Phi|\le \sqrt{-k}$.

In case $k>0$, so that the metric has positive scalar curvature
everywhere, then the only solutions have $\Phi=0$.  Then the
Seiberg--Witten equations become the anti-self-dual equations for $A$,
and usual techniques show that when the gauge group is $U(1)$, $A$
must be a flat connection, and when $X^4$ is simply connected, then up
to gauge $A$ must be the trivial connection.  Therefore, for manifolds
with positive scalar curvature, the only solution is the trivial
solution $(A,\Phi)=(0,0)$.

Furthermore, even when the manifold does not admit a metric of positive
scalar curvature, we still have the bound $|\Phi|\le \max\{-R\}$.
So when the manifold $X^4$ is compact, since the scalar curvature must
be bounded, we have that when $(A,\Phi)$ is a solution of the Seiberg--Witten
equations, $|\Phi|$ must also be bounded.  It is possible
to prove that for solutions, the connection $A$, modulo gauge, is also
bounded.  This shows that the moduli space of solutions is compact.
The fact that the moduli space is compact is one of the most important
reasons why the Seiberg--Witten equations are easier to work with than
the anti-self-dual equations.

We can try to define Seiberg--Witten invariants in analogy to
Donaldson invariants.  To do this we need to be certain that the
moduli space is always a compact manifold, and that when two different
metrics are used, $g_0$ and $g_1$, a generic path $g_t$ between them
provides a cobordism between the two moduli spaces.\footnote{It is
sometimes useful to instead perturb the first equation slightly using
a self-dual two-form, instead of perturbing the metric, since this
would require redefining the spinors, and allows removing the trivial
solution when it is not generic.  This is a technical point and not
crucial to our story.}

We've mentioned that these moduli spaces are compact, and as long as
the gauge group acts freely, we can avoid singularities if $b_2^+>1$, as
before.

When the expected dimension (\ref{eqn:swdim}) is zero, we can count
the number of solutions (with appropriate sign and multiplicity) and
we will get an invariant if $b_2^+>1$, for the same reason that this idea
works for Donaldson invariants.  When the expected dimension is
not zero, we can consider the moduli space of solutions as it sits
inside the space of fields modulo gauge, and compute what homology class
it represents, as in the situation with Donaldson theory.  In this way, we
get Seiberg--Witten invariants $SW(w)$ for each $\spin^c(4)$ structure $W$, and
thus, for each cohomology class $w=c_1(\det(W))\in H^2(X^4)$.

It is conjectured that if $b_2^+>1$, the Seiberg--Witten invariants
are only non-zero when the dimension (\ref{eqn:swdim}) is zero.  This
statement is true for the many known cases.  This condition, called
{\em Seiberg--Witten simple type}, is supposed to be equivalent to
the simple type condition for Donaldson invariants.

\section{Donaldson = Seiberg--Witten?}
The duality given by Seiberg and Witten's work is more detailed than saying
the theories are in some vague sense ``equivalent''.  The duality also predicts
particular formulas that relate Donaldson invariants to Seiberg--Witten
invariants.

Recall that (assuming $b_2^+>1$ and simple type) the Donaldson
invariants can be written as a series which can be factored as
\begin{equation}
D=e^{I/2}\left(r_1e^{K_1}+\dots+r_me^{K_m}\right)
\label{eqn:dseries2}
\end{equation}
where $I$ is the intersection form, $r_1, \dots, r_m$ are rational numbers,
and $K_1, \dots, K_m$ are elements in $H^2(X^4)$.

Recall that (assuming $b_2^+>1$ and Seiberg--Witten simple type)
the Seiberg--Witten invariants assign to each class $w\in H^2(X^4)$ an
integer $SW(w)\in \zz$ counting the number of solutions to the Seiberg--Witten
equations with sign.

I can now present the relationship between Donaldson invariants and
Seiberg--Witten invariants.  According to the (somewhat non-rigorous)
argument by Seiberg and Witten\cite{W4,MW}, the classes $K_i$
that appear in (\ref{eqn:dseries2}) are the classes $w$ for which
$SW(w)\not=0$, and the rational coefficient $r_i$ is equal to $SW(w)$,
up to a factor:
\begin{equation}
r_i=2^{\frac{1}{4}(18+14b_1+18b_2^+-4b_2^-)}SW(K_i)
\label{eqn:dsw}
\end{equation}
The power of two in front is a kind of renormalization factor, and
although the form of the power came out of the theory, the actual
coefficients were discovered by plugging in particular known examples.
The formula also holds up in many other known cases, so
many people are confident that the formula is in general true.

A rigorous proof of this result is still lacking, however.  Seiberg
and Witten's work do not constitute a proof, since there are a number
of non-rigorous arguments, ranging from concluding that no other
factors arise from integration in the $u$-plane, to the whole notion
of functional integration (which is still not founded on rigorous
mathematics, even today).  The conjectured relationship
(\ref{eqn:dsw}) turns out to work in the many cases where the
Donaldson invariants and the Seiberg--Witten invariants are both
known.  This ``empirical'' evidence may be convincing, but for
mathematicians concerned with calculating these invariants, the lack
of a rigorous proof is problematic.

In 1995, Victor Pidstrigach and Andrei Tyurin\cite{PT} proposed a
program to prove the relationship (\ref{eqn:dsw}) between the
Donaldson invariants and the Seiberg--Witten invariants.  Their
approach is to consider a theory that contains both the scalar field
$\Phi$ and the non-abelian gauge group $SO(3)$ (which is basically
$SU(2)$, except it identifies $+I$ with $-I$).  The theory they
examine is analogous to the Seiberg--Witten equations (\ref{eqn:sw}),
though slightly more complicated.

The moduli space of solutions to these equations behaves similarly to
the moduli spaces for $SU(2)$ instantons in Donaldson theory, but the
behavior of the singularities is more intricate.  Pidstrigach and
Tyurin claim that there are two kinds of singularities that can occur:
those that appear because of solutions to the anti-self-dual equation
$*F=-F$, and those that appear because of solutions to the Seiberg--Witten
equations (\ref{eqn:sw}).  The moduli space of $SO(3)$ monopoles, then,
is a cobordism between the moduli space in Donaldson theory and the
moduli space in Seiberg--Witten theory.  This would be helpful in proving
equation (\ref{eqn:dsw}).

Carrying out this program involves a great deal of difficult
mathematics, and this mathematics is being developed by Paul Feehan
and Thomas Leness in a series of
papers.\cite{FL1,FL2,F1,F2,FL3,FL4,FL5} The difficulties associated
with working with the Pidstrigach--Tyurin theory are the difficulties
with the Donaldson theory combined with the difficulties of the
Seiberg--Witten equations, so these papers involve delicate functional
analysis.  Unlike the Seiberg--Witten equations, there is no
compactness result, and the gauge group is non-abelian.  The
analytical details are still being developed by Feehan and Leness.
Meanwhile, with what they have accomplished so far, Feehan and Leness
have proved Witten's conjectured relationship (\ref{eqn:dsw}) between
Donaldson invariants and Seiberg--Witten invariants for a large class
of manifolds, up to a certain number of terms.\cite{FL3,FL4}  Given the
impressive work so far, it is reasonable to hope that this program
will eventually prove the equivalence of the Donaldson invariants and
the Seiberg--Witten invariants.

This Pidstrigach--Tyurin--Feehan--Leness program does not follow the
Seiberg--Witten S duality approach.  It might be instructive to
investigate if there is a way of phrasing this program in terms of S
duality.  If so, this might open a new way of thinking about dualities
in physics.

Even if this program does not illuminate S duality, physicists will
still benefit from this Pidstrigach--Tyurin--Feehan--Leness program.
The fact that the expected relationships between Donaldson invariants
and Seiberg--Witten invariants do work out may be an indication that
the results from Seiberg--Witten theory are true and dependable when
applied to other theories.

\section{Seiberg--Witten invariants, K\"ahler geometry, and Riemannian geometry}
The Seiberg--Witten equations might be studied independently of whether or
not they relate to the Donaldson invariants, since to a topologist, instantons
were not an end anyway, but merely a means to an end.  So it is possible
to try to find ways to study the Seiberg--Witten invariants and see what it
has to say about four-dimensional manifolds, even without linking
Seiberg--Witten invariants to Donaldson invariants.

In the case where $X^4$ carries a K\"ahler metric, the Donaldson
theory simplified considerably.  As it turns out, the Seiberg--Witten
theory simplifies even more dramatically.\cite{Mor}  When a manifold carries a
K\"ahler metric, there is a canonical class $K_X\in H^2(X)$.  In this
case, it is possible to rewrite the Seiberg--Witten equations as
equations involving complex holomorphic sections, and explicitly
derive the solutions to the Seiberg--Witten equations.  When this is
done, we see that the Seiberg--Witten invariants on $K_X$ have the
values $SW(K_X)=1$, $SW(-K_X)=\pm 1$, and all other Seiberg--Witten
invariants are zero.\footnote{There is a formula that
determines whether $SW(-K_X)$ is $1$ or $-1$, but that would be
distracting at this point.\cite{Mor}}

Among the first papers that used Seiberg--Witten theory was the proof
by Kronheimer and Mrowka\cite{KM2} of the Thom conjecture, which
states that if an embedded surface $\Sigma^2\subset \cc P^2$
represents a class in $H_2(\cc P^2)\cong \zz$, the genus of $\Sigma^2$
must be at least $(d-1)(d-2)/2$, where $d$ is integer labeling the
classes in $H_2(\cc P^2)$.  The formula $(d-1)(d-2)/2$ is interesting,
because that is exactly the genus of $\Sigma^2$ in the case where
$\Sigma^2$ is algebraic.  More generally, surfaces inside K\"ahler
manifolds that are algebraic have the least possible genus for their
homology class.\cite{MST} Besides answering an important question
relating topology and algebraic geometry, what was particularly
striking was how short the paper was compared to many papers that used
Donaldson theory to prove various kinds of results.  In other words,
Seiberg--Witten theory was easier than Donaldson theory.

Even when the manifold contains a symplectic form $\omega$ that is not
necessarily K\"ahler, Taubes showed that $SW([\omega])$ is
non-zero,\cite{T1} and even related it to counting $J$-holomorphic
curves in $X^4$ (the Gromov--Witten invariants in symplectic
geometry).\cite{T2} This has led to new ways of thinking about
symplectic geometry, $J$-holomorphic curves, and its relations to
contact geometry.

The applications to Riemannian geometry were also extensive and
intriguing.  As we saw above, when the manifold $X^4$ has a metric
with positive scalar curvature, there is only one solution, and this
can be perturbed away if $b_2^+>1$.  In other words, when the $X^4$
admits a metric of positive scalar curvature and $b_2^+>1$, then
$SW(w)=0$ for all $w\in H^2(X^4)$.

Therefore, such manifolds cannot be K\"ahler or even symplectic.
Claude LeBrun\cite{L} used arguments related to this observation to
compute the Yamabe invariants for certain K\"ahler manifolds, and
found a large class of four-dimensional manifolds that do not admit
Einstein metrics.

The idea of pulling apart four-dimensional manifolds along necks has
been more successful for Seiberg--Witten theory than with Donaldson
theory, partly because of compactness.  It is possible to define
a Floer-like homology (Seiberg--Witten--Floer homology) using solutions
to these equations on $Y^3\times\re$.\cite{Mar,Wan,MarWan,Fro,MOY,Iga,Nic}

When $Y^3$ has positive scalar curvature (for instance, $Y^3$ is a sphere
$S^3$), we get the same kind of results as in Donaldson theory, that
is, the Seiberg--Witten invariants vanish on $X^4=X_+\#X_-$ when
$b_2^+(X_+)$ and $b_2^+(X_-)$ are both positive.  Therefore, if
$X_1$ and $X_2$ have $b_2^+>1$, then $X_1\# X_2$ cannot be symplectic.

The relations of Seiberg--Witten solutions to
$J$-holomorphic curves mentioned above allow interpretations that
have implications to symplectic and contact topology, and also
suggest ways to compute Seiberg--Witten invariants (see the next
section).

There are many other situations where it can be proved that the
Seiberg--Witten invariants are zero, and whenever this happens and
$b_2^+>1$, we can be assured that the manifold is not symplectic and
therefore, not K\"ahler.  One recent example is Scott Baldridge's
work, which shows that if a manifold $X^4$ has an effective $S^1$
action with a fixed point, and $b_2^+>1$, then all the Seiberg--Witten
invariants vanish (and therefore cannot be symplectic).\cite{Bal} In
particular, the subject of symplectic manifolds with $S^1$ action has
been simplified dramatically.

\section{Using Seiberg--Witten invariants to distinguish manifolds}
The most direct use of a new invariant is to use it to distinguish
manifolds.  We now know of many examples of two manifolds that are
homeomorphic (in particular have the same homology, cohomology,
intersection form) that are not diffeomorphic, because their Seiberg--Witten
invariants are different.

Along these lines, R. Fintushel and R. Stern\cite{FS2} gave an
infinite collection of smooth manifolds homeomorphic to the K3
manifold but with different Seiberg--Witten invariants.  These were
obtained by taking a knot or link in $S^3$, and removing a small
neighborhood of the knot or link from $S^3$, then taking the resulting
manifold and forming the cartesian product with $S^1$.  The resulting
four-dimensional manifold has a $T^3$ boundary.  We then take a
K3 manifold and remove a neighborhood of a particular $T^2$, and this
gives us a four-dimensional manifold with $T^3$ boundary.  Then glue
the two manifolds together along this boundary in a particular way.

It turns out that these have the same intersection matrix as K3, so
by Freedman's work\cite{Fr}, they are all homeomorphic.  But
the Seiberg--Witten invariants are essentially the coefficients of the
Alexander polynomial of the original knot or link.  Taking different
knots or links gives different manifolds that are homeomorphic
but have different Seiberg--Witten invariants, and so are not diffeomorphic.

More generally, there is a great deal of work that relates Seiberg--Witten
Floer homology to the Alexander polynomial for links, and various kinds
of topological torsion.\cite{MT,Hut,HL}

Computing Seiberg--Witten invariants, though less hopeless than
computing Donaldson invariants, is still not necessarily easy, and it
is not clear whether or not there will eventually be a general
technique to compute them.  Work in this direction is indicated by
Peter Ozsvath and Zoltan Szab\'o, who have developed a kind of
Seiberg--Witten-like invariant that they conjecture is equal to the
Seiberg--Witten invariants, but is more combinatorial in nature and
is easier to calculate.\cite{OS1,OS2,OS3}

\section{The eleven-eighths conjecture}
The eleven-eighths conjecture is about manifolds whose intersection matrix is
broken into a certain number of $H$'s and a certain number of $E_8$'s.
The conjecture says that the number of $H$'s must be at least $3/2$ the
number of $E_8$'s, and when this is written in terms of $b_2$ and $\sigma$,
the result is $b_2\ge \frac{11}{8}|\sigma|$.

The most dramatic progress so far in proving the eleven-eighths conjecture is
the work of M. Furuta\cite{Fur}, showing that as long as we have at
least one $E_8$, the number of $H$'s must be at least one larger than
the number of $E_8$'s.  This statement can be expressed as $b_2 \ge
\frac{10}{8}|\sigma|+2$.  See Figure~\ref{fig:geography}.

This was proved by looking at the solutions to the Seiberg--Witten
equations as zeros of a non-linear operator, and approximating the
linear part of the operator by a finite-dimensional operator operating
on the first several eigenspaces.  By adding in the non-linear part,
it is possible to construct a finite-dimensional approximation to the
Seiberg--Witten operator.  The overall constant gauge symmetries
provide a symmetry in finite dimensions that give rise to equivariant
maps of spheres.  By classic $K$-theoretic results on equivariant maps
of spheres, Furuta obtains his result.

Furuta, Kametani and Matsue furthermore prove that if there are
four $E_8$'s, then there must be at least 6 $H$'s.\cite{FKM}

\section{The Future}
The Seiberg--Witten equations have been more than a way to distinguish
four-dimensional manifolds.  Apart from their relation to physics, there
are intriguing relationships to symplectic geometry, scalar curvature,
the Alexander polynomial for links, Reidemeister torsion, and so on.
It is possible that Seiberg--Witten theory is a part of a bigger picture
that unifies these concepts.  Seiberg--Witten Floer homology for
three-dimensional manifolds have similar relationships to those subjects,
but in three-dimensional topology there is already the program of
W. Thurston, that seeks to understand three-dimensional manifolds as
combinations, along spheres and incompressible tori, of geometrically
uniform three-dimensional manifolds (so that their geometry is one of
eight geometries).  The fact that Seiberg--Witten theory gives trivial
invariants when we split along spheres and to some extent along tori of
a certain type, and when the manifolds have positive scalar curvature,
might suggest that the kind of decomposition of Thurston might be
natural for Seiberg--Witten theory also, but no relationships have yet
been found.

In other words, there are a number of intriguing relationships between
the Seiberg--Witten equations and other ideas, and if these are more
than coincidental, we can look forward to many fruitful synergies that
result from understanding how these subjects are related.

The relationship to physics is perhaps the most direct one, since the
Seiberg--Witten equations came directly from physics.  It is
gratifying to see more mathematicians and physicists working together
because of these and other influences of physics on mathematics
(mirror symmetry, Yang--Baxter, Monstrous Moonshine, the Penrose
inequality, etc.).  It is possible to suggest that this collaboration
has much further to go, since mathematicians have still not found a
way to understand much of quantum field theory in ways that have sound
mathematical footing, and much of the mathematical work with Seiberg--Witten
theory uses the equations on their own terms, instead of looking to the
original supersymmetry theories.  Perhaps this is because the problems
involved in making functional integrals (for instance) completely rigorous
are considered too difficult, and certainly risky for those in a
publish-or-perish environment.  But if and when mathematicians
find a solid mathematical foundation for the arguments involved in
the work of Seiberg and Witten, there are bound to be many new
developments in both mathematics and physics.

Witten seems to hope for this, when in his conclusion on his work on
Supersymmetry and Morse theory\cite{W1} he writes:
\begin{quote}
It is not at all clear whether supersymmetry plays a role in nature.  But if it
does, this is a field in which mathematical input may make a significant
contribution to physics.
\end{quote}

In section 2 of his work relating Donaldson theory to a certain SUSY
theory,\cite{W2} Witten gets a little more explicit:
\begin{quote}
In this section, we will see what can be obtained by formal
manipulations of Feynman path integrals.  Of course, a rigorous
framework for four dimensional quantum gauge theory has not yet been
developed to a sufficient extent to justify all of our considerations.
Perhaps the connection we will uncover between quantum field theory
and Donaldson theory may serve to broaden the interest in constructive
field theory, or even stimulate the development of new approaches to
that subject.
\end{quote}

If Witten were the sort of person to say, ``I told you so,'' he would
have strong justification for doing so, in light of the Seiberg--Witten
equations.  In one of his papers with Moore on the relationship
between the Seiberg--Witten equations and Donaldson theory,\cite{MW} Witten
more modestly concludes:
\begin{quote}
In this paper, we have obtained a more comprehensive understanding of
the relation between the Donaldson invariants and the physics of
$N=2$ supersymmetric Yang--Mills theory.  In particular, we have
explained the role of the $u$-plane in Donaldson theory more
thoroughly than had been done before, both for $b_2^+=1$ and for hypothetical
manifolds of $b_2^+>1$ that are not of simple type.  We hope that in the
process the power of the quantum field theory approach to Donaldson theory
and the rationale for the role of modular functions in Donaldson theory
have become clearer.
\end{quote}

For mathematicians, the lesson is clear: never underestimate the
importance of physics is solving mathematical problems, and perhaps
effort invested in solving problems in physics will reap rewards in
mathematics down the line.

For physicists, the application of duality to an area of mathematics
may argue for the importance and validity of supersymmetry and
duality.  We might view four-dimensional topology as an experimental
apparatus.  Since mathematicians are finding that the Seiberg--Witten
invariants are really related to the Donaldson invariants in ways that
Seiberg and Witten predicted, then this lends credence to the idea
that duality, derived from mathematical manipulations that are not
always rigorous, really does work, and gives one hope that at least
supersymmetric gauge theories will one day be on firm ontological
foundation, or at least be proved as mathematically consistent as the
rest of mathematics.

Inasmuch as many important theories in physics have led to unexpected
powerful developments in mathematics, perhaps we have some inductive evidence
that the reverse is true: if a theory leads to unexpected powerful
developments in mathematics, the theory may be an important one in physics.
We have just seen that supersymmetric gauge theories lead to the kind of
development in mathematics that one might associate with a good physical
theory.

This may also be an indication that there is something deeper going
on, where a more general mathematical theory explains what why we
should expect supersymmetry to be relevant to four-dimensional
topology.  This more general theory may, in turn, lay the foundation for
new physical theories.

\section*{Acknowledgements}

Thanks to the many physicists, especially Michael Peskin, who, even
though I am a mathematician, have spent time to make the physics they
study clear and interesting to me.  I hope I have made the mathematics
I study clear and interesting to physicists in return.


\begin{thebibliography}{99}
\bibitem{NS} C. Nash and S. Sen, {\em Topology and Geometry for physicists},
Academic Press, 1983.
\bibitem{Mu} J. Munkres, {\em Elements of Algebraic Topology}, Addison
Wesley, 1984.
\bibitem{B} W. Boone, The Word problem, {\em Annal. of Math.} {\bf
70} (1959), pp. 207--265.
\bibitem{BHP} W. Boone and W. Haken and V. Poenaru, On
recursively unsolvable problems in topology and their classification,
{\em Contributions to Math. Logic} (Colloquium, Hannover, 1966),
North--Holland, 1968, pp. 37--74.
\bibitem{Ko} A. Kosinski, {\em Differential Manifolds}, Academic Press,
1993.
\bibitem{Ra} A. Ranicki, {\em Algebraic and Geometric Surgery}, Oxford
Univ. Press, 2002.
\bibitem{Ki} R. Kirby, {\em The Topology of 4-manifolds}, Springer,
Lecture notes in Mathematics no. 1374, 1989.
\bibitem{Fr} M. Freedman, The Topology of four-dimensional
manifolds, {\em Journ. Differ. Geom.}, {\bf 17} (1982), pp. 357--453.
\bibitem{FQ} M. Freedman and F. Quinn, {\em Topology of 4-manifolds},
Princeton Univ. Press, Princeton Math. Series no. 39, 1990.
\bibitem{D1} S. Donaldson, An application of gauge theory to the topology
of four-manifolds, {\em Jour. Differ. Geom.}, {\bf 18} (1983), pp. 269--316.
\bibitem{D2} S. Donaldson, The orientation of Yang--Mills moduli spaces
and 4-manifold topology, {\em Jour. Differ. Geom.}, {\bf 26} (1987), pp. 397.
\bibitem{FU} D. Freed and K. Uhlenbeck, {\em Instantons and four manifolds},
Springer 1984.
\bibitem{DK} S. Donaldson and P. Kronheimer, {\em The geometry of four-manifolds}, Oxford Univ. Press, 1990.
\bibitem{D3} S. Donaldson, Polynomial invariants for smooth 4-manifolds,
{\em Topology}, {\bf 29} (1990), pp. 257--315.
\bibitem{FM} R. Friedman and J. Morgan, {\em Smooth four-manifolds and
complex surfaces}, Springer, 1994.
\bibitem{MM} J. Morgan and T. Mrowka, A note on Donaldson's polynomial
invariants, {\em Int. Math. Research Notices}, {\bf 10} (1992), pp. 223--230.
\bibitem{PS} M. Peskin and D. Schroeder, {\em An Introduction to Quantum
Field Theory}, Addison Wesley, 1995.
\bibitem{Cole} S. Coleman, {\em Aspects of Symmetry}, Cambridge Univ. Press, 1985.
\bibitem{ADHM} M. Atiyah and V. Drinfeld and N. Hitchin and Y. Manin,
Construction of instantons, {\em Physics Letters}, {\bf 65A}, pp. 185--187.
\bibitem{AW} M. Atiyah and R. Ward,  Instantons and Algebraic geometry,
{\em Comm. Math. Phys.}, {\bf 55} (1977), pp. 117--124.
\bibitem{U} K. Uhlenbeck, Removable singularities in Yang--Mills fields.
{\em Communications in Mathematical Physics}, {\bf 83}, (1982), pp. 11--29.
\bibitem{KM1} P. Kronheimer and T. Mrowka,
Embedded surfaces and the structure of Donaldson's polynomial invariants
{\em Jour. Differ. Geom.} {\bf 41} (1995), pp. 573--734.
\bibitem{FS1} R. Fintushel and R. Stern,
Donaldson invariants of 4-manifolds with simple type,
{\em Journal of Differ. Geom.} {\bf 42} (1995), pp. 577--633.
\bibitem{Fl1} A. Floer, {\em An instanton invariant for three manifolds}, 
{\em Comm. Math. Phys.}, {\bf 118} (1988), pp. 215--240.
\bibitem{At1} M.F. Atiyah, New invariants of three and four
dimensional manifolds, {\em Symposium on the mathematical heritage of
Hermann Weyl}, R. Wells, et al., eds., Univ. of North Carolina, May
1987.
\bibitem{At2} M.F. Atiyah, {\em The Geometry and Physics of Knots},
Cambridge Univ. Press, 1990.
\bibitem{W1} E. Witten, Supersymmetry and Morse theory,
{\em Jour. Differ. Geom.},  {\bf 17} (1982), pp. 661--692.
\bibitem{W2} E. Witten, Topological Quantum Field Theory, {\em Comm. Math.
Phys.}, {\bf 117} (1988), pp. 353--386.
\bibitem{AH} L. Alvarez-Gaum\'e and S. F. Hassan, Introduction to
$S$-Duality in $N=2$ supersymmetric gauge theories, {\em Fortschritte
Phys.} {\bf 45} (1997), pp. 159--236.
\bibitem{W3} E. Witten, Supersymmetric Yang--Mills theory on a
four-manifold, {\em Jour. Math. Phys.}, {\bf 35} (1994), pp. 5101--5135.
\bibitem{S0} N. Seiberg, Supersymmetry and Non-perturbative Beta
functions, {\em Physics Letters} {\bf B206} (1988), pp. 75--80.
\bibitem{S1} N. Seiberg, Exact results on the space of vacua of
Four dimensional SUSY gauge theories, {\em Phys.Rev.} {\bf D49} (1994),
pp. 6857-6863, hep-th/9402044.
\bibitem{S2} N. Seiberg, The power of duality---exact results in
4D SUSY field theory, {\em Int'l. Journal of Modern Physics A}, {\bf 16} (2001)
pp. 4365--4376.
\bibitem{SW1} N. Seiberg and E. Witten, Electric-magnetic duality,
monopole condensation, and confinement in $N=2$ supersymmetric Yang--Mills
theory, {\em Nucl. Phys.} {\bf B426} (1994) pp. 19-52; Erratum-ibid.
{\bf B430} (1994), pp. 485-486, hep-th/9407087.
\bibitem{MO} C. Montonen and D. Olive, Magnetic monopoles as gauge particles?,
{\em Phys. Lett.} {\bf B72} (1977), pp. 117.
\bibitem{WO} E. Witten and D. Olive, Supersymmetry algebras that include
topological charges, {\em Phys. Lett.} {\bf B78} (1978), pp. 97--101.
\bibitem{SW2} N. Seiberg and E. Witten, Monopoles, Duality, and Chiral
Symmetry Breaking in $N=2$ supersymmetric QCD, {\em Nucl. Phys.},
{\bf B431} (1994), pp. 484-550, hep-th/9408099.
\bibitem{W4}E. Witten, Monopoles and Four-manifolds, {\em Math. Res. Letters},
{\bf 1} (1994), pp. 769--796.
\bibitem{MW} G. Moore and E. Witten, Integration over the $u$-plane
in Donaldson theory, {\em Adv. Theor. Math. Phys.}, {\bf 1} (1997) no. 2,
pp. 298--387.,  hep-th/9709193.
\bibitem{PT} V. Pistrigach and A. Tyurin, {\em Localisation of the Doanldson's
invariants along Seiberg--Witten classes}, dg-ga/9507004.
\bibitem{FL1} P.M.N. Feehan and T.G. Leness, $PU(2)$ monopoles. I:
Regularity, compactness and transversality, {\em J. Differential
Geom.} {\bf 49} (1998), pp. 265-410, arXiv:dg-ga/9710032.
\bibitem{FL2} P.M.N. Feehan and T.G. Leness, $PU(2)$ monopoles and
relations between four-manifold invariants, {\em Topology Appl.} {\bf 88}
(1998), pp. 111-145, arXiv:dg-ga/9709022.
\bibitem{F1} P.M.N. Feehan, Generic metrics, irreducible rank-one $PU(2)$
monopoles, and transversality, {\em Comm. Anal. Geom.} {\bf 8} (2000),
pp. 905-967, arXiv:math.DG/9809001.
\bibitem{F2} P.M.N. Feehan, Critical-exponent Sobolev norms and the
slice theorem for the quotient space of connections, {\em Pac. J. Math.}
{\bf 200} (2001), pp. 71-118, arXiv:dg-ga/9711004.
\bibitem{FL3} P.M.N. Feehan and T.G. Leness, $PU(2)$ monopoles and links
of top-level Seiberg-Witten moduli spaces, {\em J. Reine Angew. Math.}
{\bf 538} (2001), to appear, arXiv:math.DG/0007190.
\bibitem{FL4} P.M.N. Feehan and T.G. Leness, $PU(2)$ monopoles. II:
Top-level Seiberg-Witten moduli spaces and Witten's conjecture in low
degrees, {\em J. Reine Angew. Math.} {\bf 538} (2001), to appear,
arXiv:dg-ga/9712005.
\bibitem{FL5} P.M.N. Feehan and T.G. Leness, $SO(3)$ monopoles,
level-one Seiberg-Witten moduli spaces, and Witten's conjecture in low
degrees, {\em Topology Appl.}, to appear, arXiv:math.DG/0106238.
\bibitem{Mor} J.W. Morgan, {\em The Seiberg--Witten equations and applications
to the topology of smooth four-manifolds}, Mathematical Notes, Princeton
Univ. Press, 1996.
\bibitem{KM2}P.B. Kronheimer and T.S. Mrowka, The Genus of Embedded
Surfaces in the Projective Plane, {\em Math. Res. Letters}, {\bf 1} (1994),
pp. 797--808.
\bibitem{MST} J. Morgan and Z. Szab\'o and C. Taubes, A product
formula for the Seiberg--Witten invariants and the generalized Thom
conjecture, {\em Journal of Differ. Geom.}, {\bf 44} (1996), pp. 706--788.
\bibitem{T1}C.H. Taubes, The Seiberg--Witten invariants and symplectic
forms, {\em Math. Res. Letters}, {\bf 1} (1994), pp. 809--822.
\bibitem{T2}C.H. Taubes, The Seiberg--Witten and Gromov invariants,
{\em Math. Res. Letters}, {\bf 2} (1995), pp. 221--238.
\bibitem{L}C. LeBrun, Four-manifolds without Einstein metrics,
{\em Math. Res. Letters}, {\bf 3} (1996), pp. 133--147.
\bibitem{Mar} M. Marcolli, Equivariant Seiberg--Witten--Floer homology,
dg-ga 9606003.
\bibitem{Wan} B.L. Wang, Seiberg--Witten--Floer theory for homology
three-spheres, unpublished.
\bibitem{MarWan} M. Marcolli and B.L. Wang,
Equivariant Seiberg-Witten Floer homology, {\em Comm. Anal. Geom.}
{\bf 9} (2001), no. 3, pp. 451--639. 
\bibitem{Fro} K. Fr\o yshov, The Seiberg--Witten equations and four-manifolds
with boundary, {\em Math. Res. Letters}, {bf 3} (1996), pp. 373--390.
\bibitem{MOY} T. Mrowka and P. Ozsvath and B. Yu, Seiberg--Witten monopoles
on Seifert fibered spaces, {\em Comm. Anal. and Geom.}, {\bf 4} (1997),
pp. 685--791.
\bibitem{Iga} K. Iga, Stanford Univ. Ph.D. Thesis, 1998.
\bibitem{Nic} L. Nicolaescu, {\em Notes on Seiberg--Witten Theory},
Amer. Math. Soc., Grad. Studies in Math. {\bf 28}, 2000.
theory.
\bibitem{Bal} S. Baldridge, Seiberg--Witten vanishing theorem for
$S^1$-manifolds with fixed points, math.GT/0201034.
\bibitem{FS2} R. Fintushel and R. Stern, Knots, links, and 4-manifolds,
{\em Inventiones Mathematicae}, {\bf 134} (1998), pp. 363--400.
\bibitem{MS}J.W. Morgan and Z. Szab\'o, Homotopy $K3$ surfaces and
mod 2 Seiberg--Witten Invariants, {\em Math. Res. Letters} {\bf 4} (1997),
pp. 17--21.
\bibitem{MT}G. Meng and C. Taubes, $SW$=Milnor Torsion, {\em Math. Res. Lett.},
{\bf 3} (1996), pp. 661--674.
\bibitem{Hut}M. Hutchings, {\em Reidemeister Torsion in generalized Morse
theory}, Harvard Univ. Ph.D. Thesis, 1998.
\bibitem{HL}M. Hutchings and Y. Lee, Circle-valued Morse theory and Reidemeister
torsion, {\em Topology}, {\bf 38} (1999), pp. 861--888.
\bibitem{OS1}P. Ozsvath and Z. Szab\'o, {\em Holomorphic disks and
three-manifold invariants: properties and applications}, preprint.
\bibitem{OS2}P. Ozsvath and Z. Szab\'o, {\em Holomorphic disks and
topological invariants for rational homology three-spheres}, preprint.
\bibitem{OS3}P. Ozsvath and Z. Szab\'o, {\em Holomorphic triangles and
invariants for smooth four-manifolds}, preprint.
\bibitem{Fur}M. Furuta, The Monopole Equations and the $11/8$ conjecture,
{\em Math. Res. Letters}, {\bf 8} (2001), pp. 279--291.
\bibitem{FKM}M. Furuta and Y. Kametani and H. Matsue, Spin 4-manifolds with signature=$-32$,
{\em Math. Res. Letters}, {\bf 8} (2001), pp. 293--301.
\end{thebibliography}
\end{document}